\documentclass[11pt,showpacs]{revtex4}
\usepackage{graphicx}
\begin{document}

\preprint{}

\title{Large photon productions in a gravitational collapsing}

\author{She-Sheng Xue}

\email{xue@icra.it}

\affiliation{ICRA and
Physics Department, University of Rome ``La Sapienza", 00185 Rome, Italy}



\begin{abstract}
We  study a possible gravitational vacuum-effect, in which vacuum-energy 
variation is due to variation of gravitational field, vacuum state gains gravitational 
energy and releases it by spontaneous photon emissions. Based on the path-integral 
representation, we present a general formulation of vacuum transition matrix 
and energy-momentum tensor of a quantum scalar field theory in curved spacetime. 
Using analytical continuation of dimensionality of the phase space,
we calculate the difference of vacuum-energy 
densities in the presence and absence of gravitational field. Using the dynamical 
equation of gravitational collapse, we
compute the rate of vacuum state gaining gravitational energy. 
Computing the transition amplitude from initial vacuum state to final
vacuum state in gravitational collapsing process, we show the 
rate and spectrum of spontaneous photon emissions for releasing gravitational energy. 
The possible connection of our study to the genuine origin of gamma ray bursts 
is discussed. We compare our idea with the Schwinger idea for Sonoluminiescence 
and contrast our scenario with the Hawking effect. 

\end{abstract}


\maketitle

\section{ Introduction}\label{introduction}

The issue of quantum field theories of elementary particles in curved spacetime 
has played a tremendously important role in understanding quantum phenomenon of 
particle creations from the vacuum, when gravitational field is present. 
In quantum field theories, the vacuum state consists of a large number of virtual 
particles, that are quantum-field fluctuations whose energy-momentum are off-shell. 
The vacuum energy (zero-point energy) is attributed to the energy-momentum of virtual particles. 
These virtual particles interact with a external gravitational field via their energy-momentum. 
Such interaction results in vacuum-energy variations and particle creations from the vacuum.   

The Hawking radiation\cite{hawking75, book} is the phenomenon of particle creations occurring around black 
hole's horizon. The quantum radiation of the Hawking type is rather general in curved spacetime\cite{xueNP2003}. 
Although, such quantum radiation is too small to be detected in the present Universe, its impact on 
theoretical understanding of particle creations due to a static gravitational field is indeed 
far reaching. On the other hand, it is important to study the phenomenon of particle creations in a non-static 
curved spacetime. In fact, the rate and spectrum of particle and antiparticle creations 
in extremely early Universe are possibly related to the CMB pattern and the large-scale structure of 
the present Universe, which are most exciting arena of theoretical and observational
physics today. Beside, in a gravitational collapse process, violent variation of gravitational field 
possibly causes a large number of particle creations, which might be account for most energetical 
events of gamma ray bursts\cite{xueshort2003}. This is the topic that we attempt to further discuss 
in this article. 

The phenomenon of particle creations from the vacuum was first studied 
by Euler, Heisenberg and Schwinger in the quantum field theory of electromagnetic dynamics
(QED) \cite{sw}. A strong external electric field, its strength is larger than $m_e^2c^3/\hbar$, greatly 
reduces the energy-mass gap ($\sim 2m_e$) of charged virtual particles in the QED vacuum, so that
virtual particles undergo a quantum tunneling process, leading to large pair-productions of 
electrons and positrons. The mechanism and phenomenon of electron and positron pair-productions 
have advocated a numerous studies both in  
experimental physics\cite{swexp} and theoretical physics\cite{swth}, as well as in 
astrophysics\cite{ruffini}. In addition, it is worthwhile to mention
that an external magnetic field can induce vacuum decay leading to spontaneous photon emissions\cite{xuephd2003}. 
The dynamics of this effect, which is completely different from the Schwinger one, is that the external
magnetic field modifies the vacuum-energy spectrum so that the 
vacuum state gains magnetic energy and becomes unstable. This effect can be possibly 
tested in laboratories and account for the phenomenon of anomalous x-ray emission 
from pulsars\cite{xrayano}. 

Casimir \cite{casimir} first considered that boundary conditions (two conducting plates) modify 
the energy spectrum of the vacuum state and vacuum-energy variation is the Casimir energy 
$\delta {\cal E}|_{\rm casimir}$. An attractive force between two plates is observed 
as the Casimir effect\cite{casimirexp}. The reasons for the Casimir effect are that the vacuum state gains 
the Casimir energy and becomes energetically unstable, quantum-field 
fluctuations result in releasing the Casimir energy. 
Recently, there is much theoretical and experimental attention on the dynamical Casimir 
effect\cite{dyacasimir}. In such an effect, boundary conditions are dynamically time-dependent, 
the Casimir energy is released by spontaneous photon emissions.     

Sonoluminiescence\cite{sololuexp,sololuth} is another most interesting vacuum effect which shows a 
flash of spontaneous photon 
emissions, when gas bubbles in water collapse, driven by a sound-wave. The vacuum-energy variation 
is due to the variation of dielectric constant, rather than the modification of boundary conditions 
as in the Casimir effect. We will have a more detailed discussion in this article\cite{swason}. 

The mystery of energetic sources generating gamma ray bursts\cite{piran,romagrb} is  
a prompt emission (seconds) of extremely huge energy output ($\sim 10^{54}$ergs for isotropic
emission) from rather compact sources ($\sim 10^8$cm) at cosmological distance 
($z\sim O(1)$). These have stimulated many studies in connection with electromagnetic 
properties of black holes\cite{ruffini,putten,koide}. Various astrophysical scenarios are
discussed in literatures\cite{piran}. It is worthwhile to mention that via the Schwinger mechanism of 
electron-positron pair productions, the ``dyadosphere''\cite{ruffini, romagrb} of photons, electrons and positrons, 
is formed during the process of gravitational collapse of a
massive star with electromagnetic structure.  

In this article, we want to study a possible gravitational vacuum-effect, in which vacuum-energy 
variation is due to the variation of gravitational field, vacuum state gains gravitational 
energy and releases it by spontaneous photon emissions. In section (\ref{general}), using path-integral 
representation, we present a general formulation of quantum scalar field theories 
in curved spacetime, where the vacuum states are defined, the transition matrix from vacuum 
to vacuum, vacuum-energy spectrum and vacuum energy-momentum tensor are obtained. 
In sections (\ref{spectrum}) and (\ref{density}), we specify two static observers respectively 
in the Schwarzschild and flat spacetime; we analyze the eigenvalues (spectrum) of the transition matrix 
and define the phase space of vacuum states; by using analytical continuation 
of dimensionality of the phase space, we obtain vacuum-energy density. In sections (\ref{scale}) 
and (\ref{computation}), we calculate vacuum-energy density and discuss why the characteristic 
energy-scale should be the ultraviolet cutoff for considering the difference of vacuum-energy 
densities in the presence and absence of gravitational field. In section (\ref{svariation}), 
we define vacuum energy with respect to two static observers and compute the difference of
vacuum energies in the presence and absence of gravitational field, which shows vacuum state gains 
gravitational energy. In section (\ref{collapsing}), we adopt a simple model describing the gravitational 
collapse of a massive shell that is infinitesimally thin, and using the dynamical equation of gravitational 
collapse, we compute the rate of vacuum state gaining gravitational energy. In section (\ref{production}), based on the transition amplitude from initial vacuum state to final
vacuum state at each step of gravitational collapsing process, we compute the rate and spectrum of spontaneous 
photon emissions for releasing gravitational energy. In sections (\ref{sololu}) and 
(\ref{hawking}), we compare our proposal for gamma ray bursts with the Schwinger proposal for 
Sonoluminiescence and contrast 
our scenario with the Hawking effect. In the final section (\ref{discussion}), we make some 
remarks on this preliminary study and the possible connection of our proposal 
to the genuine origin of gamma ray bursts.    
 
\section{\it General formulation.}\label{general}

We assume that the structure of spacetime is described by the pseudo-Riemannian metric $g_{\mu\nu}$
associated with the line element
\begin{equation}
ds^2=g_{\mu\nu}dx^\mu dx^\nu,\hskip0.5cm \mu,\nu=0,1,2,3,
\label{line}
\end{equation}
and the spacetime point is coordinated by $x=(x^0,x^i)=(t,\vec x)$.
The special geometrical symmetries of the spacetime $\cal S$ are described by using 
Killing vectors $\xi^\mu$, which are solutions of Killing's equation
\begin{equation}
{\cal L}_\xi g_{\mu\nu}(x)=0,\hskip0.5cm \xi_{\mu;\nu}+\xi_{\nu;\mu}=0,
\label{killing}
\end{equation}
where ${\cal L}_\xi$ is the Lie derivative along the vector field $\xi^\mu$, orthogonal to the 
spacelike hypersurface $\Sigma_t$ ($t=$constant) of the spacetime ${\cal S}$. 
A static observer ${\cal O}$ is at rest in this hypersurface $\Sigma_t$. 

We consider that at the initial time ($t_{\rm in}= -\delta t/2$), the spacetime is asymptotically flat, 
described by asymptotically flat geometry $\bar g_{\mu\nu}$ in Eq.(\ref{line}); 
while at the final time ($t_{\rm out}= +\delta t/2$), the spacetime is curved and stationary, 
described by a non-trivial geometry $g_{\mu\nu}$ in Eq.(\ref{line}).  
The characteristic time-scale $\delta t$ of such variation of spacetime geometry
is supposed to be much larger than the characteristic time-scale (e.g., $1/m_e$) of a 
quantum-field transition, i.e., $\delta t\gg 1$ in the unit of the quantum time-scale.  
We attempt to study quantum-field fluctuations interacting with the variation of spacetime geometry.
A specific model for such variation of spacetime geometry and quantum-field transition
will be presented in due course.

In order to clearly illustrate the physics content,
we first consider a complex scalar field $\phi$ in curved spacetime. 
The simplest coordinate-invariant action is given by ($\hbar=c=G=k=1$)
\begin{equation}
S = {1\over2}\int
d^4x\sqrt{-g}\Big[g^{\mu\nu}\phi_{,\mu}\phi^*_{,\nu}+(m^2+\xi {\cal R})
\phi\phi^*\Big],
\label{action}
\end{equation}
where $m$ is particle mass and ${\cal R}$ the Riemann scalar. 
The quantum scalar field $\phi$ can be in principle expressed in terms of
a complete and orthogonal basis of quantum-field states $u_k(x)$:
\begin{equation}
\phi(x)=\sum_k\Big(a_k u_k(x)+
a^\dagger_ku^*_k(x)\Big),\hskip0.1cm \left [a_k,a^\dagger_{k'}\right ]
=\delta_{k,k'}
\label{de}
\end{equation}
where $a^\dagger_k$ and $a_k$ are creation and annihilation operators
of the $k$-th quantum-field state $u_k(x)$. This quantum field state obeys the 
following equation of motion,
\begin{equation}
(\Delta_x + m^2+\xi {\cal R})u_k(x)=0,
\label{eq}
\end{equation}
and appropriate boundary conditions for selected values of $k$. In Eq.(\ref{eq}),
$\Delta_x$ is the Laplacian operator in curved spacetime:
\begin{equation}
\Delta_x = {1\over\sqrt{-g}}\partial_\mu\big[\sqrt{-g}g^{\mu\nu}\partial_\nu\big].
\label{laplace}
\end{equation}
In Eq.(\ref{de}), we assume that $u_k(x)$ are positive energy ($\omega$) states, 
satisfying
\begin{equation}
{\cal L}_\xi u_k(x)=-i\omega u_k(x), \hskip0.5cm \omega >0,
\label{killingv}
\end{equation} 
with respect to the timelike Killing vector field $\xi^\mu$ (\ref{killing}) associated to 
the static observer ${\cal O}$ rest in hypersurface $\Sigma_t$ ($t=$constant) 
of the spacetime ${\cal S}$. 

At the initial time ($t_{\rm in}= -\delta t/2$), the spacetime is approximately flat, 
quantum-field states $u_k(x)$ (\ref{eq})
are asymptotically free states $\bar u_k(x)$, obeying Eq.(\ref{eq}) for the asymptotically free 
geometry $\bar g_{\mu\nu}$. Then, the quantum scalar field $\phi(x)$ is an 
asymptotically free field in the hypersurface $\Sigma_{-\delta t/2}$ of the asymptotically
free spacetime $\bar {\cal S}$:
\begin{equation}
\phi_{\rm in}(x)=\sum_k\Big(\bar a_k \bar u_k(x)+
\bar a^\dagger_k\bar u^*_k(x)\Big),\hskip0.1cm \left [\bar a_k,\bar a^\dagger_{k'}\right ]
=\delta_{k,k'}
\label{de1}
\end{equation}
where $\vec x\in \Sigma_{-\delta t/2}$, $\bar a^\dagger_k$ and $\bar a_k$ are creation and 
annihilation operators of the $k$-th asymptotically free quantum-field state $\bar u_k(x)$. 
Corresponding Lie derivative along the Killing vector (\ref{killing}) is $\partial_t$, 
positive energy states are $\bar u_k(x)$, satisfying Eq.(\ref{killingv}). 
Then we may construct the standard Minkowski space quantum vacuum state 
$|\bar 0,{\rm in}\rangle$:
\begin{equation}
\bar a_k |\bar 0,{\rm in} \rangle=0,\hskip0.5cm \langle \bar 0,{\rm in}|\bar a_k^\dagger=0.
\label{vd1}
\end{equation}
$|\bar 0,{\rm in} \rangle$ is an initial quantum vacuum state at the initial time 
($t_{\rm in}= -\delta t/2$) with respect to the static observer ${\cal O}$ 
rest in hypersurface $\Sigma_{-\delta t/2}$ of the spacetime $\bar {\cal S}$.

At the final time ($t_{\rm out}= +\delta t/2$), the spacetime ${\cal S}$ is curved, 
described by a stationary non-trivial geometry $g_{\mu\nu}$. We assume that 
quantum-field states $u_k(x)$ (\ref{eq}) are asymptotical states $\tilde u_k(x)$, 
obeying Eq.(\ref{eq}) for the stationary geometry $g_{\mu\nu}$. In the hypersurface 
$\Sigma_{+\delta t/2}$ of the spacetime ${\cal S}$,
the asymptotical quantum scalar field $\phi_{\rm out}$ is expressed in terms of $\tilde u_k(x)$:
\begin{equation}
\phi_{\rm out}(x)=\sum_k\Big(\tilde a_k \tilde u_k(x)+
\tilde a^\dagger_k \tilde u^*_k(x)\Big),\hskip0.1cm \left 
[\tilde a_k,\tilde a^\dagger_{k'}\right ]=\delta_{k,k'}
\label{de2}
\end{equation}
where $\vec x\in \Sigma_{+\delta t/2}$, $\tilde a^\dagger_k$ and $\tilde a_k$ are creation and 
annihilation operators of the $k$-th quantum-field state $\tilde u_k(x)$. 
Corresponding Lie derivative along the Killing vector is $\xi^\mu$ (\ref{killing}), positive energy 
states are $\tilde u_k(x)$ satisfying Eq.(\ref{killingv}). Then we may construct the
quantum vacuum state $|\tilde 0,{\rm out}\rangle$:
\begin{equation}
\tilde a_k |\tilde 0,{\rm out}\rangle=0,\hskip0.5cm 
\langle \tilde 0,{\rm out}|\tilde a_k^\dagger=0,
\label{vd}
\end{equation}
in curved spacetime.
$|\tilde 0,{\rm out} \rangle$ is an final quantum vacuum state at the final time 
($t_{\rm out}= +\delta t/2$) with respect to the same static observer ${\cal O}$
rest in hypersurface $\Sigma_{+\delta t/2}$ of the spacetime ${\cal S}$.
 
It is worthwhile to note that $\phi_{\rm out}(x)$($\{\tilde u_k(x)\}$) are not asymptotically free field(states),
instead they are asymptotical field(states) in the presence of external and stationary gravitational field,
so that the final vacuum state $|\tilde 0,{\rm out}\rangle$ (\ref{vd}) is different from 
the initial vacuum state $|\bar 0,{\rm in}\rangle$ (\ref{vd1}). 
Such a difference is not only a unitary phase.
The final vacuum state $|\tilde 0,{\rm out}\rangle$ (\ref{vd}), may not necessarily be measured as devoid of particles, in contrast to the initial vacuum state $|\bar 0,{\rm in}\rangle$ 
defined by Eq.(\ref{vd1}) relating to the asymptotically free field $\phi_{\rm in}(x)$.
In fact, as will be shown, the final vacuum state $|\tilde 0,{\rm out}\rangle$ is a quantum-field 
state of particle and antiparticle creations upon the initial vacuum state $|\bar 0,{\rm in}\rangle$. 
This indicates gravitational 
field interacting with quantum-field fluctuations of positive and negative energy states 
of the initial vacuum $|\bar 0,{\rm in} \rangle$, and the quantum scalar field evolves throughout 
intermediate quantum-field states $\phi(\vec x,t)$ (\ref{de}) for $-\delta t/2 <t<+\delta t/2$. 
We speculate that this evolution is adiabatic for $\delta t$ being much larger than quantum-field transition time.  

To deal with all possible intermediate states, represented by $\phi(x)$ or $u_k(x)$ 
in Eq.(\ref{de}) for $-\delta t/2 <t<+\delta t/2$, we use path-integral representation 
to study the transition amplitude between the initial vacuum state and final vacuum state: 
\begin{equation}
\langle \tilde 0,{\rm out}|\bar 0,{\rm in}\rangle=
\int [{\cal D}\phi{\cal D}\phi^*]\exp(iS),
\label{action2}
\end{equation}
where 
\begin{equation}
\int [{\cal D}\phi{\cal D}\phi^*]=\Pi_{-\delta t/2 <t<+\delta t/2}\Pi_{\vec x\in\Sigma_t}
\int [d\phi(\vec x,t)\phi^*(\vec x,t)].
\label{measure}
\end{equation}
The intermediate quantum-field states contributions to the transition amplitude (\ref{action2}) 
can be formally path-integrated,
\begin{equation}
\langle \tilde 0,{\rm out}|\bar 0,{\rm in}\rangle 
= {\det}^{-1}\left({\cal M}\right),\hskip0.2cm
{\cal M}\! =\!\Delta_x + m^2+\xi {\cal R}.
\label{m}
\end{equation}
This result clearly depends on the initial vacuum state $\phi_{\rm in}$ (\ref{de1}) and final vacuum 
state $\phi_{\rm out}$ (\ref{de2}), which are not explicitly written. The effective action $S_{\rm eff}$ 
is defined as
\begin{equation}
S_{\rm eff}=-i\ln \langle \tilde 0,{\rm out}|\bar 0,{\rm in}\rangle,
\label{action1}
\end{equation}
which relates to the phase of the $S$-matrix transition from the initial vacuum state 
$|\bar 0,{\rm in}\rangle$ to the final vacuum state $|\tilde 0,{\rm out}\rangle$. 
The averaged energy-momentum tensor 
$\langle T_{\mu\nu}\rangle$ of the quantum-field vacuum is given by:
\begin{equation}
\langle T^{\mu\nu}(x)\rangle = 
{2\over\sqrt{-g}}{\delta S_{\rm eff}\over \delta
g_{\mu\nu}(x)}.
\label{e1t}
\end{equation}
Main efforts are calculations of the effective action (\ref{action1}) and 
energy-momentum tensor (\ref{e1t}) in this article. 

In order to evaluate the path-integral (\ref{action2}) over all intermediate quantum-field states $\phi(x)$ 
(\ref{de}), it is convenient to introduce operators $\hat X_\mu$ and $\hat K_\mu$ 
defined on the states $|x\rangle$ and $|k\rangle$:
\begin{equation}
\hat X_\mu|x\rangle=x_\mu|x\rangle;\hskip0.3cm \hat K_\mu|k\rangle=k_\mu|k\rangle.
\label{xk}
\end{equation}
They enjoy the canonical communication,
\begin{equation}
[\hat X_\mu,\hat K_\nu]=-ig_{\mu\nu}.
\label{cxk}
\end{equation}
The states $|x\rangle$ and $|k\rangle$ satisfy:
\begin{eqnarray}
\langle x| x'\rangle &=&\delta(x-x'),\hskip0.5cm 
\int dx|x\rangle \langle x|=1\nonumber\\
\langle k| k'\rangle &=& 2\pi\delta(k-k'),\hskip0.3cm  
\int dk|k\rangle \langle k|=1,
\label{basis}
\end{eqnarray}
and intermediate quantum-field state $u_k(x)$ can be represented as
\begin{equation}
u_k(x)=\langle x|k\rangle.
\label{uxk}
\end{equation}
Using these matrix notations, we write the operator ${\cal M}(\hat X,\hat K)$ (\ref{m}) 
as a hermitian matrix 
\begin{eqnarray}
{\cal M}^{x,x'}_{k,k'}\! &=& \!\langle k|x\rangle\langle x|{\cal M}(\hat X,\hat K)| x'
\rangle\langle x'| k' \rangle\nonumber\\
\! &=&\! u^*_k(x){\cal M}(x,\hat K)u_{k'}(x')\delta(x-x'),\nonumber\\
\! &=&\! u^*_k(x){\cal M}(x,\hat K)u_{k'}(x')\delta_{kk'}\delta(x-x'),
\label{mxk}
\end{eqnarray}
where 
\begin{equation}
\langle x|{\cal M}(\hat X,\hat K)| x'\rangle={\cal M}(x,\hat K)\delta(x-x'),
\end{equation} 
is diagonal in the coordinate space $\{x\}$.
In the representation of $\{u_k(x)\}$, the hermitian matrix ${{\cal M}^{x,x'}_{k,k'}}$ 
is also diagonal in the energy-momentum space $\{ k\}$, since $\{u_k(x)\}$ 
are eigenstates of the operator ${\cal M}$ (\ref{eq}). By normalizing the 
quantum field $\phi$, we define 
the normalized diagonal element of the matrix (\ref{mxk}) as
\begin{equation}
\lambda^2_k(x) \equiv {1\over |u_k(x)|^2} u^*_k(x){\cal M}(x,\hat K)u_{k}(x),
\label{lambdat}
\end{equation}
and formally compute the effective action $S_{\rm eff}$ given in Eqs.(\ref{action1},\ref{m}):
\begin{eqnarray}
iS_{\rm eff} &=& -\ln \langle \tilde 0,{\rm out}|\bar 0,{\rm in}\rangle iS_{\rm eff}\nonumber\\
&=&-\int \sqrt{-g}{d^4xd^4k\over (2\pi)^4}
\ln\lambda^2_k(x)|_{\rm out}\nonumber\\
&-&\left( -\int \sqrt{-g}{d^4xd^4k\over (2\pi)^4}
\ln\lambda^2_k(x)|_{\rm in}\right),
\label{zr}
\end{eqnarray}
where the $\{\lambda^2_k\}_{\rm out}$ and $\{\lambda^2_k\}_{\rm in}$ are the diagonal elements 
(\ref{lambdat}). The $\{\lambda^2_k\}_{\rm out}$ is in terms of the final vacuum state and 
geometry  $g_{\mu\nu}$, whereas the $\{\lambda^2_k\}_{\rm in}$ is in terms of the initial 
vacuum state $\phi_{\rm in}$ and geometry $\bar g_{\mu\nu}$:
\begin{eqnarray}
\lambda^2_k|_{\rm out} &=& {1\over |\tilde u_k(x)|^2}\tilde u^*_k(x){\cal M}(x,\hat K)\tilde u_{k}(x),
\label{lambdatout}\\
\lambda^2_k|_{\rm in} &=& {1\over |\bar u_k(x)|^2}\bar u^*_k(x){\cal M}(x,\hat K)\bar u_{k}(x).
\label{lambdatin}
\end{eqnarray}
The operator $\lambda^2_k(x)$ (\ref{lambdat}) and the number of quantum-field states 
$\int \sqrt{-g}d^4xd^4k/(2\pi)^4$ are invariant in
arbitrary coordinate systems, later 
is the Liouville theorem for the phase-space invariance.

By using Eqs.(\ref{e1t}) and (\ref{zr}), the variation $\langle T_{\mu\nu}(x)\rangle_{\rm diff}$
of averaged energy-momentum tensor in the time interval $\delta t$ is given by the difference:
\begin{equation}
\langle T^{\mu\nu}(x)\rangle_{\rm diff} = \langle T^{\mu\nu}(x)\rangle_{\rm out}-
\langle T^{\mu\nu}(x)\rangle_{\rm in},
\label{tem}
\end{equation}
where $\langle T^{\mu\nu}(x)\rangle_{\rm out}$ is the averaged energy-momentum tensor 
computed by the variation of geometry $g_{\mu\nu}$ (\ref{e1t}), corresponding to the final quantum vacuum state $\phi_{\rm out}$, whereas $\langle T^{\mu\nu}(x)\rangle_{\rm in}$ 
is the averaged energy-momentum tensors computed by the variation of geometry 
$\bar g_{\mu\nu}$ (\ref{e1t}), corresponding to the initial quantum vacuum state 
$\phi_{\rm in}$. 

In general, using the definition (\ref{zr}), we formally calculate
the averaged energy-momentum tensor $\langle T_{\mu\nu}(x)\rangle$ as,
\begin{equation}
\langle T^{\mu\nu}(x)\rangle = \langle T^{\mu\nu}(x)\rangle^{(1)}
+\langle T^{\mu\nu}(x)\rangle^{(2)},
\label{em}
\end{equation}
where
\begin{eqnarray}
\langle T^{\mu\nu}(x)\rangle^{(1)} &=& i g^{\mu\nu}(x)\int{d^4k\over (2\pi)^4}
\ln (\lambda^2_k(x))\label{e1'}\\
\langle T^{\mu\nu}(x)\rangle^{(2)} &=& 2i\int{d^4k\over (2\pi)^4}{\delta
\ln (\lambda^2_k(x))\over\delta g_{\mu\nu}(x)}.
\label{e1'2}
\end{eqnarray}
In these equations
\begin{eqnarray}
&&{\delta \sqrt{-g(x)}\over \delta g_{\mu\nu}(y)}={1\over2}\sqrt{-g(x)}
g^{\mu\nu}(x)\delta^4(x-y);\nonumber\\
&&\int d^4x\sqrt{-g} \delta^4(x-y) f(x)= f(y),
\label{fq}
\end{eqnarray}
for an arbitrary function $f(x)$. The calculations of energy-momentum tensor (\ref{e1'},\ref{e1'2})
are main tasks in the following two sections.
 
\section{Vacuum Energy-Spectrum.}\label{spectrum}

In order to illustrate physical idea in a mathematically tractable way, we model a collapsing massive star
as an infinitesimally thin and spherical shell. This massive shell separates the spacetime into
two regions: (i) internal region $\bar {\cal S}$ described by the flat geometry $g_{\mu\nu}=(1,-1,-1,-1)$,
\begin{equation}
ds^2\!=\!dt^2_0\!-\!dr^2\!-\!r^2d\Omega,\hskip0.3cm r<R;
\label{sg1}
\end{equation}
and (ii) the external spacetime ${\cal S}$ described by the stationary Schwarzschild geometry,
\begin{eqnarray}
ds^2\!&=&\!g_{tt}dt^2\!+\!g_{rr}dr^2\!-\!r^2d\Omega,\hskip0.3cm r>R\nonumber\\
g_{tt}\!&=&-(g_{rr})^{-1}=g\!\equiv\! (1\!-\!{2M\over r}),
\label{sg}
\end{eqnarray}
where $r,\theta,\phi$ are spherical-polar coordinates, $d\Omega=d\theta^2+\sin^2\theta d\phi^2$,
$t$ and $t_0$ are the Schwarzschild-like coordinates in the external and internal 
spacetime respectively. $R$ indicates the radial position of the shell. $M$ is the total 
mass-energy of the shell.

At the initial time $t_{\rm in}=-\delta t/2$, the shell radius is $R$ and a static observer ${\cal O}$ 
is located at $\vec x (R-0^+,\Omega)\in\Sigma_{-\delta t/2}$ in the internal spacetime 
$\bar {\cal S}$. His four velocity $u_\mu$ and Killing vector $\xi_\mu$ are given by, 
\begin{equation}
u_\mu = (1,0,0,0),\hskip0.5cm
\xi_\mu  = (1,0,0,0).
\label{o1}
\end{equation}
The initial quantum field is $\phi_{\rm in}$ (\ref{de1}) and quantum 
vacuum state is $|\bar 0,{\rm in}\rangle$ (\ref{vd1}). After $\delta t$, the shell gravitationally collapses
and its radial position $R$ moves inwards to $R-\delta R$. At the final time $t_{\rm out}=+\delta t/2$,
the same static observer ${\cal O}$ turns out to be in the external 
spacetime ${\cal S}$. His four velocity $u_\mu$ and Killing vector $\xi_\mu$ are then given by, 
\begin{equation}
u_\mu = (g^{1/2}_{tt}(r),0,0,0),\hskip0.5cm
\xi_\mu  = (g_{tt}(r),0,0,0).
\label{o}
\end{equation}
The final quantum field is $\phi_{\rm out}$ (\ref{de2}) and quantum 
vacuum state is $|\tilde 0,{\rm out}\rangle$ (\ref{vd}). In the following, we respectively compute 
the vacuum-energy spectrum of the initial and final vacuum states. 

In the external spacetime, the Riemann scalar ${\cal R}=0$ and the Laplacian operator 
(\ref{laplace}) is given by:
\begin{eqnarray}
\Delta_x &=& g^{tt}{\partial^2\over \partial t^2}+{1\over r^2}{\partial\over \partial r} 
r^2g^{rr}{\partial\over \partial r}
+{\hat L^2\over r^2}\nonumber\\
&=&g^{tt}{\partial^2\over \partial t^2} +g^{rr}\left({\partial^2\over \partial r^2}
+{2\over r}{\partial\over \partial r}\right)\nonumber\\
&-&{2M\over r^2}{\partial\over \partial r}
+{\hat L^2\over r^2},
\label{laplace1}
\end{eqnarray}
where $\hat L^2$ is the angular momentum operator, 
$g^{tt}= (g_{tt})^{-1}$ and $g^{rr}= (g_{rr})^{-1}$.

The appropriate basis of quantum field states 
(\ref{de2}) is chosen as 
\begin{equation}
\tilde u_k(x)= \langle t,r,\theta,\phi|\omega,k_r,l,m\rangle =R_{l\omega}(r)Y_{lm}
(\theta,\phi)e^{-i\omega t},
\label{sphere}
\end{equation} 
where $Y_{lm}(\theta, \phi)$ is the standard spherical harmonic function: 
$\hat L^2Y_{lm}(\theta, \phi)=l(l+1)Y_{lm}(\theta, \phi)$ and 
$k$ indicates a set of quantum numbers $(\omega,k_r,l,m)$. $\omega$ 
is the energy-spectrum and the radial momentum $k_r$ will be defined soon. From Eq.(\ref{eq}),
the radial function $R_{l\omega}(r)$ obeys the following differential equation for $r>R>2M$,
\begin{equation}
\left[ g^{tt}\omega^2+g^{rr}\hat k_r^2+i{2M\over r^2}\hat k_r-V_l\right]R_{l\omega}(r)=0,
\label{eq2}
\end{equation}
where the hermitian radial momentum operator,
\begin{equation}
\hat k_r={1\over ir}{\partial\over\partial r}r,\hskip0.3cm \hat k_r^2
=-\left({\partial^2\over\partial r^2}+{2\over r}{\partial\over\partial r}\right),
\label{pr}
\end{equation}
and potential
\begin{equation}
V_l={l(l+1)\over r^2}+{2M\over r^3}+m^2.
\label{p}
\end{equation}
Eq.(\ref{eq2}) is exactly equivalent to the Regge and Wheeler equation. The radial 
function $R_{l\omega}(r)$ is an orthogonal and complete basis,
asymptotically behaves as the Hankel function $h_{l\omega}(r)$ for $r\gg R>2M$.
The matrix operator ${\cal M}$ in Eq.(\ref{lambdat}) is given by,
\begin{equation}
{\cal M}(r,\hat k_r)= g^{tt}{\partial^2\over \partial t^2}-g^{rr}\hat k_r^2
-i{2M\over r^2}\hat k_r+V_l.
\label{matrix}
\end{equation}
Eqs.(\ref{sphere}-\ref{matrix}) define a complex eigen-value problem to find the energy 
spectrum of the final vacuum state in an external gravitation field. The imaginary part of the  
operator ${\cal M}(r,\hat k_r)$ (\ref{matrix}) results in the quantum radiation of Hawking
type in curved spacetime, as discussed in ref.\cite{xueNP2003}.
 
The diagonal matrix $\lambda^2_k$ (\ref{lambdat}) is given by:
\begin{equation}
\lambda^2_k(x)|_{\rm out}=-g^{tt}\omega^2-g^{rr}k_r^2+ V_l-i{2M\over r^2}k_r,
\label{ipro0}
\end{equation}
where we define the values of ``radial momentum'' $k_r$ and $k_r^2$ 
of the quantum field state $\tilde u_{k}(x)$:
\begin{eqnarray}
k_r &\equiv&  {\tilde u^*_k \hat k_r \tilde u_k \over |\tilde u_k|^2}
={R^*_{\omega l} \hat k_r R_{\omega l} \over |R_{\omega l}|^2};\label{kr0}\\
k^2_r &\equiv& {\tilde u^*_k \hat k_r^2 \tilde u_k \over |\tilde u_k|^2}
={R^*_{\omega l} \hat k_r^2 R_{\omega l} \over |R_{\omega l}|^2}.
\label{kr20}
\end{eqnarray}

In the internal spacetime $\bar {\cal S}$, the Laplacian operator 
(\ref{laplace}) is given by:
\begin{equation}
\Delta_x = {\partial^2\over \partial t^2_0}-{1\over r^2}{\partial\over \partial r} 
r^2{\partial\over \partial r}
+{\hat L^2\over r^2}.
\label{laplace1f}
\end{equation}
The appropriate basis of quantum field states 
(\ref{de1}) is chosen as 
\begin{equation}
\bar u_k(x)= \langle t_0,r,\theta,\phi|\omega_0,k_{r0},l,m\rangle =\bar R_{l\omega_0}(r)Y_{lm}
(\theta,\phi)e^{-i\omega_0 t_0}.
\label{spheref}
\end{equation} 
The radial function $\bar R_{l\omega_0}(r)$ obeys the following differential equation,
\begin{equation}
\left[ \omega_0^2- \hat k_r^2-{l(l+1)\over r^2}-m^2\right]\bar R_{l\omega_0}(r)=0.
\label{eq2f}
\end{equation}
The radial function $\bar R_{l\omega_0}(r)$ forms a orthogonal and complete basis.
The matrix operator ${\cal M}$ in Eq.(\ref{lambdat}) is given by,
\begin{equation}
{\cal M}(r,\hat k_r)= {\partial^2\over \partial t_0^2}+\hat k_r^2
+{l(l+1)\over r^2}+ m^2.
\label{matrixf}
\end{equation}
 The diagonal matrix $\lambda^2_k$ (\ref{lambdat}) is given by:
\begin{equation}
\lambda^2_k(x)|_{\rm in}=-\omega_0^2+ k_{r0}^2+ {l(l+1)\over r^2}+ m^2,
\label{ipro0f}
\end{equation}
where we define the values of ``radial momentum'' $k_{r0}$ and $ k_{r0}^2$ 
of the quantum field state $\bar u_{k}(x)$:
\begin{eqnarray}
k_{r0} &\equiv&  {\bar u^*_k \hat k_r \bar u_k \over |\bar u_k|^2}
={\bar R^*_{\omega_0 l} \hat k_r \bar R_{\omega_0 l} \over |\bar R_{\omega_0 l}|^2};\label{kr01}\\
k_{r0}^2 &\equiv& {\bar u^*_k \hat k_r^2 \bar u_k \over |\bar u_k|^2}
={\bar R^*_{\omega_0 l} \hat k_r^2 \bar R_{\omega_0 l} \over |\bar R_{\omega_0 l}|^2}.
\label{kr201}
\end{eqnarray}
For $m=0$ in Eq.(ref{eq2f}), 
$\bar R_{\omega_0 l}=2\omega_0 j_l(\omega_0 r)$, $j_l(\omega_0 r)$ is the spherical 
Bessel function. As a particular case, we adopt the spherically symmetric solution $l=0$ 
and $m=0$ in the differential equation Eq.(\ref{eq2f}), such that the spherically symmetric 
solution $\bar R_{0\omega_0}(r)\sim e^{ik_rr}/ r$ is the eigenstate of the radial momentum 
operators $\hat k_r$ and $\hat k^2_r$ (\ref{pr}). The eigenvalues $k_{r0}$ (\ref{kr01}) and 
$k_{r0}^2$ (\ref{kr201}) are related to radial momentum of the spherical quantum field $(l=0)$. 
This indicates that
the values of ``radial momentum'' $k_r$ (\ref{kr0}) and $k_r^2$ (\ref{kr20}) 
are consistent with the radial momentum of quantum field states.

\section{Vacuum Energy Density}\label{density}

Armed with the energy-momentum tensor $\langle T^{\mu\nu}(x)\rangle$ (\ref{em}) and the 
energy-spectrum $\lambda^2_k(x)|_{\rm out}$ (\ref{ipro0}) in the external spacetime ${\cal S}$, 
we are ready to compute the energy-density,
\begin{equation}
\langle T^t_t\rangle_{\rm out} =g_{tt}\langle T^{tt}\rangle_{\rm out}.
\label{enden}
\end{equation}
Given the volume of spherical shell $4\pi r^2 drdt$, 
we have the number of quantum field states in the energy-momentum phase-space,
\begin{equation}
\int{d^4k\over (2\pi)^4} = \int {d^2k_\perp\over (2\pi)^2}\int {d\omega dk_r\over (2\pi)^2}={1\over 4\pi r^2}\sum_{l,m}\int {d\omega dk_r\over (2\pi)^2},
\label{kv}
\end{equation}
where $k_r$ is the ``radial momentum'' defined in Eq.(\ref{kr0}) and $\vec k_\perp$ are the 
transverse momenta, perpendicular to the radial direction.
 
Starting with the first part $\langle T^{\mu\nu}(x)\rangle^{(1)}$ of the energy-momentum 
tensor (\ref{e1'}), we compute the vacuum energy-density (\ref{enden}) as:
\begin{equation}
\langle T^t_t\rangle^{(1)} = {i\over 4\pi r^2}\sum_{l,m}\int {d\omega dk_r\over (2\pi)^2} 
\ln (\lambda^2_k).
\label{t}
\end{equation} 
Using the identity:
\begin{equation}
\ln{a\over b}=\int_0^\infty {ds\over s}\left(e^{is(b+i\epsilon)}-e^{is(a+i\epsilon)}\right).
\label{id}
\end{equation}
we are able to write the vacuum energy-density (\ref{t}),
\begin{eqnarray}
\langle T^t_t\rangle^{(1)} &=&  -{i\over 4\pi r^2}
\sum_{l,m}\int {d\omega dk_r\over (2\pi)^2}
\int_0^\infty {ds\over s}e^{is(\lambda^2_k+i\epsilon)}\nonumber\\
&+&(\lambda^2_k\rightarrow 1).
\label{zr10}
\end{eqnarray}
The second term indicated by $(\lambda^2_k\rightarrow 1)$ is 
the same as the first term with $\lambda^2_k\rightarrow 1$, 
and this term is a constant. 
The logarithmatic function in Eq.(\ref{t}) is represented by 
an $s$-integration in Eq.(\ref{zr10}) and infrared convergence at $s\rightarrow 0$ is 
insured by $i\epsilon$ prescription ($\epsilon\rightarrow 0$). 

Introducing the variable $\xi=-i\omega$ (the Wick rotation) and the integral representation,
\begin{equation}
\int^\infty_{-\infty}{d\xi\over (2\pi)}
e^{-i\beta \xi^2}={1\over 2\sqrt{i\pi \beta}},
\label{angu}
\end{equation}
for ${\rm Im}(\beta)<0$, we express Eq.(\ref{zr10}) as,
\begin{eqnarray}
\langle T^t_t\rangle^{(1)} &=& -{1 \over 8\pi r^2}\sum_{l,m}
\int{dk_r\over (2\pi)}{1\over \sqrt{-i\pi g^{tt}}}\int_0^\infty
{ds\over s^{3/2}}
e^{is( \lambda^2_k + i\epsilon)},\nonumber\\
&+&(\lambda^2_k\rightarrow 1)
\label{seff}
\end{eqnarray}
where and henceforth, $\lambda^2_k$ is given by Eq.(\ref{ipro0}) without the 
$\omega^2$-term.

In order to compute the integration over ``$s$'' in Eq.(\ref{seff}), we introduce a complex
variable $z=-1/2+\delta$ ($|\delta| \rightarrow 0 $), and use the following integral
representation of the $\Gamma(z)$-function by an analytical continuation 
for ${\rm Im}(\alpha) >0$:
\begin{equation}
\int_0^\infty e^{i\alpha s}s^{z-1}ds=(-i\alpha)^{-z}\Gamma(z).
\label{int}
\end{equation}
This analytical continuation is equivalent to analytical continuation of 
dimensionality
of the momentum-space $\int{d\omega\over (2\pi)}$ in Eq.(\ref{angu}).
In the neighborhood of singularity, where $|\delta| \rightarrow 0$ and 
$z\rightarrow -1/2$ in Eq.(\ref{int}), we have 
\begin{equation}
\Gamma(z)=-2\sqrt{\pi},\hskip0.2cm
\alpha^{-z}=\sqrt{gk_r^2+ V_l-i{2M\over r^2}k_r}.
\label{ana}
\end{equation}
Substituting Eqs.(\ref{int},\ref{ana}) into Eq.(\ref{seff}), we obtain the vacuum energy-density:
\begin{equation}
\langle T^t_t\rangle^{(1)}={g^{1/2}\over 4\pi r^2}\sum_{l,m}
\int{dk_r\over (2\pi)}\sqrt{gk_r^2\!+\! V_l\!-\!i{2M\over r^2}k_r},
\label{real2}
\end{equation}
where the ``radial momentum'' integration $\int dk_r$ integrates from ``0'' to a ultra violate cutoff $\Lambda$.

Now we turn to the computations of the second part $\langle T^{\mu\nu}(x)\rangle^{(2)}$ 
of the energy-momentum tensor (\ref{e1'2}). Using Eqs.(\ref{e1'2},\ref{enden}) and exchanging
momentum-integration and metric-variation, we write
\begin{equation}
\langle T^t_t\rangle^{(2)} =2g_{tt}{\delta\over\delta g_{tt}(x)}\left[i\int{d^4k\over (2\pi)^4}
\ln (\lambda^2_k(x)) \right],
\label{enden2}
\end{equation} 
where the bracket $[\cdot\cdot\cdot]$ was computed, as shown in Eq.(\ref{seff}).
In addition, we have $\delta (g^{tt}g_{tt})=0$, $g^{tt}\delta g_{tt} =-g_{tt}\delta g^{tt}$ and
\begin{equation}
g_{tt}{\delta\over\delta g_{tt}}{1\over \sqrt{g^{tt}}}
=-g^{tt}{\delta\over\delta g^{tt}}{1\over \sqrt{g^{tt}}}={1\over 2\sqrt{g^{tt}}}.
\label{deri}
\end{equation} 
Using Eq.(\ref{seff}) and relationship (\ref{deri}), we take metric-derivative in Eq.(\ref{enden2}) and 
obtain the result:
\begin{equation}
\langle T^t_t\rangle^{(2)} = \langle T^t_t\rangle^{(1)}.
\label{seff2}
\end{equation}
Thus the vacuum-energy density in the external spacetime ${\cal S}$ is
\begin{eqnarray}
\langle T^t_t\rangle_{\rm out} &=&\langle T^t_t\rangle^{(1)}+ \langle T^t_t\rangle^{(2)}\nonumber\\
&=&2{g^{1/2}\over 4\pi r^2}\sum_{l,m}
\int{dk_r\over (2\pi)}\sqrt{gk_r^2\!+\! V_l\!-\!i{2M\over r^2}k_r}.
\label{real3}
\end{eqnarray}
This is the vacuum energy-density corresponding to the final vacuum state $\phi_{\rm out}$.

In the internal flat spacetime $\bar {\cal S}$, 
the vacuum-energy density (\ref{real3}) is reduced to,
\begin{equation}
\langle T^t_t\rangle_{\rm in} = 2{1\over 4\pi r^2}\sum_{l,m}
\int{dk_{r0}\over (2\pi)}\sqrt{k_{r0}^2+ {l(l+1)\over r^2}+ m^2}.
\label{real5}
\end{equation}
This is the vacuum energy-density corresponding to the final vacuum state $\phi_{\rm in}$.
We adopt the transverse momenta $\vec k_\perp$ to replace the quantum numbers 
$(l,m)$ of angular momenta:
\begin{equation}
|\vec k_\perp |^2\simeq {l(l+1)\over r^2},\hskip 0.5cm  
\int {d^2k_\perp\over (2\pi)^2}={1\over 4\pi r^2}\sum_{l,m}.
\label{ka}
\end{equation}
Eq.(\ref{real5}) can be written as
\begin{equation}
\langle T^t_t\rangle_{\rm in}=
\int{d^3k\over (2\pi)^3}\sqrt{k^2+ m^2},
\label{real6}
\end{equation}
where $k^2=k_{r0}^2+|\vec k_\perp |^2$, $d^3k=dk_{r0} d^2k_\perp$ and integration 
range $[-\Lambda,\Lambda]^3$.
Eq.(\ref{real6}) is consistent with the vacuum-energy density 
for two components of quantum scalar field.

Comparing the vacuum-energy density $\langle T^t_t\rangle_{\rm out}$ (\ref{real3}) in the 
Schwarzschild geometry with the vacuum-energy density $\langle T^t_t\rangle_{\rm in}$ 
(\ref{real5}) in the flat geometry, 
we find the vacuum-energy of the quantum-field fluctuations (virtual particles) of the vacuum
couples to the gravitational field in the following aspects: (i) an addition term $2M/ r^3$ in 
the potential $V_l$ (\ref{p}) and an imaginary term $i2Mk_r/r^2$; (ii) 
the ``radial momentum'' $k_r$ and $k_r^2$ are modified from Eqs.(\ref{kr0},\ref{kr20}) 
to Eqs.(\ref{kr01},\ref{kr201}); (iii) modification $gk_r^2\rightarrow k_{r0}^2$. 
This shows the radial modes described by $k_r$ 
directly interact with gravitational field. While, the transverse 
momenta $\vec k_\perp$ in Eq.(\ref{real3}) are the same as its counterpart in Eq.(\ref{real5}), showing the
transverse modes described by $\vec k_\perp$ do not directly interact with gravitational field.  
In following sections, we attempt to calculate the difference of vacuum-energy densities, 
\begin{equation}
\langle T^t_t(x)\rangle_{\rm diff} = \langle T^t_t(x)\rangle_{\rm out}-
\langle T^t_t(x)\rangle_{\rm in},
\label{tem1}
\end{equation}
where $\langle T^t_t(x)\rangle_{\rm out}$ is in terms of the final quantum vacuum state 
$\phi_{\rm out}$ and geometry $g_{\mu\nu}$; whereas $\langle T^t_t(x)\rangle_{\rm in}$ 
is in terms of the final quantum vacuum state $\phi_{\rm in}$ and geometry $\bar g_{\mu\nu}$.

\section{Three energy scales $\Lambda, m$ and $r^{-1}$}\label{scale}

The vacuum-energy densities Eqs.(\ref{real3}) and (\ref{real5})
depend on the three very different energy scales: $\Lambda$, $m$, $r^{-1}$ (
$\Lambda\gg m\gg r^{-1}$). The scale $M$ is considered the same as the scale $r^{-1}$. 
One would expect that final results of vacuum-energy densities $\langle T^t_t\rangle_{\rm out}$ (\ref{real3}) and 
$\langle T^t_t\rangle_{\rm in}$ (\ref{real5}) should be in terms of finite terms: $r^{-4}, mr^{-3}, m^2r^{-2}$ 
and $m^4$, since all ``divergent terms'' containing $\Lambda$ should be removed away.
In what follows, we argue that $\Lambda$ is a physical cutoff of its own right for calculating 
the difference (\ref{tem1}) of vacuum energies in the presence and absence of 
gravitational field;
``divergent terms'' containing $\Lambda$ should not be simply removed away.

We know that a quantum field theory in the flat spacetime, although its characteristic 
energy-scale is the measured mass $m$ of particles, has involved high-energy modes in the 
computations of virtual particle contributions (loops in Feynman diagrams) to 
quantum corrections. The ultraviolet divergent terms arise from the contributions 
of these high-energy virtual particles. A ultraviolet cutoff $\Lambda$ or another 
method is introduced to regularize these ultraviolet divergent terms, which
are then consistently removed away by renormalizing quantum fields and parameters of theory.
Only relevant contributions of virtual particles, whose wave-lengths are the order 
of $m^{-1}$, are taken into account. This is due to the fact that high-energy virtual 
particles characterized by the cutoff $\Lambda$ must not contribute to a physical process 
characterized by the energy scale $m$, that is much smaller than $\Lambda$. 
In a sensible quantum field theory, theoretical results corresponding to experimental 
low-energy measurements, must be independent of the ultraviolet cutoff. 
The ultraviolet divergent terms can be consistently removed, if and only if the 
number of renormalized fields and parameters (for instance, coupling and mass)
is exactly equal to the number of different types of ultraviolet 
divergent terms. This is normally guaranteed by internal symmetries of a renormalizable 
quantum field theory.

The vacuum energy of a quantum field theory is a ``divergent constant'', 
since it is resulted from virtual particles whose energy range from 0 to 
the cutoff $\Lambda$. In the flat spacetime, only the energy-difference of 
quantum states can be measured, as a consequence the vacuum energy can be 
discarded, by normal ordering of quantum fields. However, in studying the 
Casimir effect, we need to calculate the difference of vacuum energies 
in presence and absence of two conducting plates, 
since the presence of two conducting plates modifies vacuum state and its energy.
This difference of vacuum energies turns out to be a measurable Casimir effect. 
The resultant Casimir energy is characterized by the 
size of the distance $L$ between two conducting plates. This is because:
\begin{itemize}
\item (i) those virtual
particles of wavelengths comparable with $L$ strongly impacted by two plates and 
their contributions to vacuum energy are modified, giving rise to the Casimir effect; 
\item (ii) whereas those virtual particles of wavelength incomparable (either much smaller 
or greater than) with $L$ are not much affected by two plates, so that their 
contributions to vacuum energy are same, and as a result these virtual particles 
have no contributions to the Casimir effect.
\end{itemize} 

The Hawking effect is certainly due to virtual particles in the vacuum interacting 
with an external static gravitational field near the horizon of a black hole. 
The reasons that this effect is characterized by the energy scale $T\sim 1/M$, which is
the gravitational potential at the size of a black hole, are the following:
\begin{itemize} 
\item (1) the dynamics of such an effect is virtual particles quantum-mechanically 
tunneling through the gravitational potential near the horizon of a black hole; 
\item (2) those 
virtual particles, whose energy is comparable with the gravitational potential around 
the horizon, have a large probability of undergoing the tunneling process, since there 
are more crossing energy-levels and virtual particles at these energy-levels 
can tunnel out of the gravitational potential into infinity;
\item (3) those virtual 
particles, whose energy is much smaller than the gravitational potential,
have a small probability of undergoing the tunneling process, since there are 
not crossing energy-levels and virtual particles at these energy-levels cannot 
tunnel out of the gravitational potential into infinity;
\item (4) those virtual particles, whose energy is much larger than the gravitational potential,
are not disturbed by the gravitational potential and remain as virtual particles 
in the vacuum.
\end{itemize}
Analogously, the quantum radiation in curved spacetime discussed in 
ref.\cite{xueNP2003} has the characteristic energy scale $\sim 1/r$, 
where is the gravitational potential that virtual particles tunnel through.    
  
We turn back to our case: we study a quantum field theory in an external 
gravitational field of a collapsing massive shell and compute the difference (\ref{tem1})
of vacuum 
energies in the presence and absence of gravitational field.
Gravitational field is a classical field of long wavelengths $O(r)$.
In curved spacetime, the vacuum energy cannot be simply be discarded,
since gravitational field couples to the vacuum energy. As we can
see the vacuum to vacuum transition matrix Eq.(\ref{ipro0}), the term 
\begin{equation}
-g^{tt}\omega^2-g^{rr}k_r^2
\label{gcvp}
\end{equation}
shows that gravitational field couples to the energy-momentum $(\omega, k_r)$ of virtual 
particles in the vacuum. High-energy modes of virtual particles have much stronger 
interactions with gravitational field than low-energy modes of virtual particles do. 
The cutoff $\Lambda$ should be a real physical cutoff, determined 
by an energy scale, where the difference (\ref{tem1}) of vacuum energies in the presence and absence of 
gravitational field vanishes. We speculate that the cutoff $\Lambda$ seems to be the Planck 
energy scale $\Lambda_p$ for the reasons: (i) virtual particles of high-energy up the 
Planck energy scale interact with gravitational field and (ii) we do not see any intermediate 
energy scale, where the difference (\ref{tem1}) of vacuum energies in the presence and 
absence of gravitational field vanishes. However,
we leave 
the ultraviolet cutoff $\Lambda$ as a parameter determined by the phenomenon of gamma 
ray bursts. We will also have a similar discussion on this point, in comparison with the  
ultraviolet cutoff $K$ introduced by Schwinger for the phenomenon of Sonoluminiescence  
in section (\ref{sololu}).   

To end this section, we note two points: (i) the difference (\ref{tem1}) of vacuum-energy 
densities $\langle T^t_t\rangle_{\rm out}$ 
and $\langle T^t_t\rangle_{\rm in}$ receives dominate contributions from high-energy modes 
of virtual particles, in particular, those modes at the ultra violate
cutoff $\Lambda$; (ii) the back action of vacuum-energy variation to gravitational field 
is not considered.

\section{Computations of vacuum energy densities}\label{computation}

We are interested in computing vacuum-energy densities Eq(\ref{real3}) in the presence 
of gravitation field and Eq.(\ref{real5}) in the absence of gravitation field. Our aim is to 
find the difference (\ref{tem1}) of these vacuum-energies. Therefore the modes of 
virtual particles that do not interact with the gravitational field should be discarded. 
As we can see from the potential term $V_l$ (\ref{p}) in Eq.(\ref{real3}), 
the radial modes described by $k_r$ and the transverse modes by angular quantum numbers 
($l,m$) or transverse momenta $\vec k_\perp $ (\ref{ka}) play very 
different roles in contributing to the vacuum-energy density. The radial 
modes of virtual particles of high energy up to the cutoff scale $\Lambda$, directly interacting 
with gravitational field, are most relevant, since they make dominate contributions to the 
difference of vacuum energies. While, the transverse modes do not directly 
interact with gravitational field, since the term $k_\perp^2=l(l+1)/r^2$ in 
(\ref{real3}) is the same as that in (\ref{real5}). 
Thus, we separate the S-wave ($l=0$) contributions from non 
S-wave ($l\not=0$) contributions to energy-momentum tensors (\ref{real3},\ref{real5}) in the 
following computations. In this section, we compute the S-wave ($l=0$) contributions 
$\langle T^t_t\rangle_{\rm in}^{l=0}$ and $\langle T^t_t\rangle_{\rm out}^{l=0}$.  
The calculations of non S-wave ($l\not=0$) contributions 
$\langle T^t_t\rangle_{\rm in}^{l\not=0}$ 
and $\langle T^t_t\rangle_{\rm out}^{l\not=0}$ will be presented in appendix A.

We first compute vacuum-energy density $\langle T^t_t\rangle_{\rm in}$ (\ref{real5}) in the 
absence of gravitational field. 
Separating the S-wave contributions $\langle T^t_t\rangle_{\rm in}^{l=0}$ from 
non S-wave contributions $\langle T^t_t\rangle_{\rm in}^{l\not=0}$, we write
Eq.(\ref{real5}) as,
\begin{equation}
\langle T^t_t\rangle_{\rm in} = \langle T^t_t\rangle_{\rm in}^{l=0}+ 
\langle T^t_t\rangle_{\rm in}^{l\not=0},
\label{sl0}
\end{equation}
where
\begin{eqnarray}
\langle T^t_t\rangle_{\rm in}^{l=0} &=& 2{1\over 4\pi r^2}
\int{dk_{r0}\over (2\pi)}\sqrt{k_{r0}^2+ m^2},
\label{l0}\\
&=&{\Gamma(-{\epsilon\over2})\over 4\pi^2 r^2}
\left(m^2\right)^{(1+\epsilon)}.
\label{l0'}
\end{eqnarray}
In Eq.(\ref{l0'}), we use the dimensional regularization of 
$\int dk_{r0}\rightarrow \int d^{(1+\epsilon)}k_{r0}$ 
in Eq.(\ref{l0}) and analytic continuation representation:
\begin{equation}
\int d^{(1+\epsilon)} k_{r0}\sqrt{k_{r0}^2+ {\cal A}^2}=\Gamma(-{\epsilon\over2})
[{\cal A}^2]^{(1+\epsilon)},
\label{fk}
\end{equation} 
where $\epsilon\rightarrow 0$ and $\Gamma(-\epsilon/2)\sim -2/\epsilon$.
The divergent term $2/\epsilon$ represents the ultraviolet cutoff $\Lambda$ 
of ``radial momentum'' $k_{r0}$. 

We turn to compute the vacuum-energy density $\langle T^t_t\rangle_{\rm out}$ 
(\ref{real3}) in 
the presence of gravitational field. Separating the S-wave contributions from
non S-wave contributions in Eq.(\ref{real3}), we have,
\begin{equation}
\langle T^t_t\rangle_{\rm out} = \langle T^t_t\rangle_{\rm out}^{l=0}+ 
\langle T^t_t\rangle_{\rm out}^{l\not=0},
\label{sgl}
\end{equation}
where
\begin{equation}
\langle T^t_t\rangle_{\rm out}^{l=0} = 2{g^{1/2}\over 4\pi r^2}
\int{dk_r\over (2\pi)}\sqrt{gk_r^2\!+\! V_{l=0}\!-\!i{2M\over r^2}k_r},
\label{gl0}
\end{equation}
where $V_{l=0}$ is $V_l(r)$ (\ref{p}) for $l=0$. 
Changing the integrating variable in Eq.(\ref{gl0}):
\begin{equation}
k_r\rightarrow k'_r=g^{1/2}k_r,\hskip0.3cm
k'_r\rightarrow \bar k_r=k'_r-i{M\over r^2g},
\label{changing}
\end{equation}
we write Eqs.(\ref{gl0}) as,
\begin{eqnarray}
\langle T^t_t\rangle_{\rm out}^{l=0} &=& {2\over 4\pi r^2}\int{d\bar k_r\over (2\pi)}
\sqrt{\bar k_r^2 + m^2 + Q^2},
\label{sgl1+}\\
&=&{\Gamma(-{\epsilon\over2})\over 4\pi^2 r^2}
\left(m^2+Q^2\right)^{(1+\epsilon)},
\label{sgl1+'}
\end{eqnarray}
where 
\begin{equation}
Q^2= {2M\over r^3} + {M^2\over gr^4}.
\label{q2}
\end{equation}
In Eq.(\ref{sgl1+'}), we use the formula (\ref{fk}) to compute the $\bar k_r$-integrations 
in $\langle T^t_t\rangle_{\rm out}^{l=0}$ (\ref{sgl1+}), 
analogously to the calculations of $\langle T^t_t\rangle_{\rm in}^{l=0}$ (\ref{l0'}). 
 
Using $\langle T^t_t\rangle_{\rm in}^{l=0}$ (\ref{l0'}) and 
$\langle T^t_t\rangle_{\rm out}^{l=0}$ (\ref{sgl1+'}) for $\epsilon\rightarrow 0$ 
up to $O(2/\epsilon)$, we find that the term $m^2$ is canceled in the difference 
$\langle T^t_t\rangle_{\rm diff}^{l=0}$ (\ref{tem1}).
To compute the $k_{r0}$- and $\bar k_r$-integrations in $\langle T^t_t\rangle_{\rm in}^{l=0}$ 
(\ref{l0}) and $\langle T^t_t\rangle_{\rm out}^{l=0}$ (\ref{sgl1+}), 
we use the formula,
\begin{eqnarray}
\int_0^{\Lambda} dx\sqrt{ax^2+ b^2} &=& {\Lambda\over 2}\sqrt{a\Lambda^2
\!+\! b^2}\label{int1}\\
&\!+\!& {b^2\over 2a^{1/2}}\ln\left({\Lambda a^{1/2}
\!+\!\sqrt{a\Lambda^2\!+\! b^2}\over b}\right).\nonumber
\end{eqnarray}
For $\Lambda\gg m\gg r^{-1}$, to the leading term ($O(\Lambda^2)$) we approximately have,
\begin{eqnarray}
\langle T^t_t\rangle_{\rm in}^{l=0} &\simeq &{1\over 4\pi r^2}\left(
{\Lambda^2\over 2\pi}\right);
\label{fl0'}\\
\langle T^t_t\rangle_{\rm out}^{l=0} &\simeq& {g\over 4\pi r^2}\left(
{\Lambda^2\over 2\pi}\right).
\label{sgl10}
\end{eqnarray}
The vacuum-energy densities Eq.(\ref{fl0'}) and Eq.(\ref{sgl10}) are mainly 
contributed from virtual particles at the ultraviolet cutoff scale $\Lambda$.

Taking into account the non S-wave contributions obtained in appendix A, we obtain 
the total vacuum-energy densities $\langle T^t_t\rangle_{\rm in}$ (\ref{sl0}) and 
$\langle T^t_t\rangle_{\rm out}$ (\ref{sgl}),
\begin{eqnarray}
\langle T^t_t\rangle_{\rm in} &\simeq & {1\over 4\pi r^2}{5\over6}\left(
{\Lambda^2\over 2\pi}\right);
\label{fl0}\\
\langle T^t_t\rangle_{\rm out} &\simeq& {g\over 4\pi r^2}{5\over6}\left(
{\Lambda^2\over 2\pi}\right),
\label{sgl1}
\end{eqnarray}
up to the leading order $O(\Lambda^2)$.

We rewrite vacuum-energy densities Eq.(\ref{fl0}) and Eq.(\ref{sgl1}) as  
\begin{eqnarray}
\langle T^t_t\rangle_{\rm in} &\simeq &{\Lambda\over 4\pi r^2dr}{5\over6}\left(
{\Lambda dr\over 2\pi}\right);
\label{fl0c}\\
\langle T^t_t\rangle_{\rm out} &\simeq & {g^{1/2}\Lambda\over 4\pi 
r^2 g^{-{1/2}}dr}{5\over6}\left(
{g^{1/2}\Lambda  g^{-{1/2}}dr\over 2\pi}\right).
\label{sgl1c}
\end{eqnarray}
Eq.(\ref{fl0c}) indicates that in the shell-volume $4\pi r^2dr$, the number of quantum-field 
states is $\left(\Lambda dr/ 2\pi \right)$ and these states carry the energy $\Lambda$. 
Eq.(\ref{sgl1c}) shows that the number of quantum-field states $\left(\Lambda dr/ 2\pi\right)$ 
is invariant, the energy of these states receives the gravitational 
red-shift $g^{1/2}(r)$ and the shell-volume $4\pi r^2dr$ receives the gravitational 
factor $g^{-{1/2}}(r)$.
 
\section{Vacuum-Energy difference and its variation}\label{svariation}

We have obtained the vacuum-energy densities $\langle T^t_t\rangle_{\rm in}$ (\ref{fl0}) 
and $\langle T^t_t\rangle_{\rm out}$ (\ref{sgl1}), and in this section we compute vacuum 
energies in the presence and absence of gravitational field. With respect to a static 
observer in spacetime (\ref{killing}), whose killing vector and four velocity
are $\xi_\mu$ and $u_\mu$, we define the vacuum energy as
\begin{equation}
{\cal E}(x) \equiv  u^\mu \langle T_{\mu\nu} \rangle u^\nu d\Sigma 
=n^\mu \langle T_{\mu\nu} \rangle d\Sigma^\nu,
\label{denergy}
\end{equation}
and
\begin{equation}
{\cal E}_{\rm total} \equiv \int_{\Sigma_t}u^\mu \langle T_{\mu\nu} \rangle u^\nu d\Sigma 
=\int_{\Sigma_t}n^\mu \langle T_{\mu\nu} \rangle d\Sigma^\nu,
\label{edefine}
\end{equation}
where $\Sigma_t$ is the hypersurface described by the equation 
$t={\rm constance}$, $d\Sigma^\nu$ its hypersurface element vector,
\begin{equation}
d\Sigma^\nu= n^\nu d\Sigma,\hskip0.5cm 
n^\nu\equiv{\xi^\nu\over\sqrt{|\xi^\alpha\xi_\alpha|}}={\xi^\nu\over \sqrt{g_{tt}}},
\label{vector}
\end{equation}
and the hypersurface element $d\Sigma=\sqrt{h}d^3x$, $h_{ij}$ is the spatial metric. 

At the initial time $(t_{\rm in}=-\delta t/2)$, with respect to the static observer 
${\cal O}$ (\ref{o1}) located in the internal flat spacetime $\bar {\cal S}$, 
the vacuum energy is given by
\begin{eqnarray}
{\cal E}_{\rm in}(r) &=& 4\pi r^2dr \langle T^t_t\rangle_{\rm in},\nonumber\\
&\simeq &{5\over6} \Big({\Lambda^2 dr\over2\pi}\Big),
\label{energy0}
\end{eqnarray}
where the vacuum-energy density $\langle T^t_t\rangle_{\rm in}$ is approximately
given by Eq.(\ref{fl0}). 
While, at the final time $(t_{\rm out}=+\delta t/2)$, 
with respect to the same static observer ${\cal O}$ (\ref{o}) located in the 
external spacetime ${\cal S}$, the vacuum energy is given by 
\begin{eqnarray}
{\cal E}_{\rm out}(r) &=& \sqrt{h}d^3x \langle T^t_t\rangle_{\rm out},\nonumber\\
&=&4\pi r^2(-g_{rr})^{1/2} dr \langle T^t_t\rangle_{\rm out},\label{birrell}\\
&\simeq & g^{1/2}(r){5\over6} \Big({\Lambda^2 dr\over2\pi}\Big),
\label{energy3}
\end{eqnarray}
where the vacuum-energy density $\langle T^t_t\rangle_{\rm out}$ is approximately given by
Eq.(\ref{sgl1}). Note that Eq.(\ref{birrell}) is the same as Eq.(6.184) in the book by 
Birrell and Davies \cite{book}.

Using Eq.(\ref{energy0}) for ${\cal E}_{\rm in}(r)$ and Eq.(\ref{energy3}) for 
${\cal E}_{\rm out}(r)$,
we approximately obtain the vacuum energy ${\cal E}_{\rm out}(r)$ 
in the presence of gravitational 
field, in terms of the vacuum energy ${\cal E}_{\rm in}(r)$ in the absence of gravitational field, 
\begin{equation}
{\cal E}_{\rm out}(r)=g^{1/2}(r){\cal E}_{\rm in}(r),
\label{energy1}
\end{equation}
which are the same as the results obtained by using the Heisenberg uncertainty relationship 
in ref.\cite{xueshort2003}. We find that the vacuum energy ${\cal E}_{\rm out}(r)$ in the presence of gravitational 
field is gravitationally red-shifted from the vacuum energy ${\cal E}_{\rm in}(r)$ in the absence of 
gravitational field. 

Corresponding to the difference $\langle T^t_t(x)\rangle_{\rm diff}$ (\ref{tem1}), 
the difference between the vacuum energy 
(\ref{energy3}) and vacuum energy (\ref{energy0}) is
\begin{equation}
\delta {\cal E}(r)\equiv {\cal E}_{\rm out}(r)- {\cal E}_{\rm in}(r) =(g^{1/2}(r)-1){\cal E}_{\rm in}(r) <0,
\label{ddem}
\end{equation}
which shows the vacuum energy gets smaller, implying that the vacuum state gains gravitational energy 
when gravitational field is turned on in the time interval $\delta t=t_{\rm out}-t_{\rm in}$. 
For $r\gg 2M$, we approximately obtain the difference 
(\ref{ddem}) of vacuum energies, 
\begin{equation}
\delta {\cal E}(r)= {\cal E}_{\rm out}(r)- {\cal E}_{\rm in}(r) 
\simeq -{M  {\cal E}_{\rm in} (r) \over r}.
\label{de0}
\end{equation}
This indicates an interacting energy due to the vacuum energy ${\cal E}_{\rm in}(r)$ 
coupling to the negative gravitational potential $-M/r$.

For the static observer ${\cal O}$ located at $r$ in the flat spacetime $\bar {\cal S}$, absolute 
value of the vacuum energy ${\cal E}_{\rm in}(r)$ (\ref{energy0}) 
is not measurable. Analogously, the static observer ${\cal O}$ located at $r$ in the curved spacetime 
${\cal S}$ is not able to measure the absolute value of the vacuum energy 
${\cal E}_{\rm out}(r)$ (\ref{energy3}). The question is then how to show 
the difference (\ref{ddem}) of vacuum energies in the presence 
and absence of a static gravitational field on the Earth. In the short letter\cite{xueshort2003}, 
author suggested to measure the Casimir effect at different altitudes above the Earth 
so as to reveal the difference of vacuum energies due to gravitational field acting on the vacuum.
Although the gravitational field is static, it varies in the radial position, the static observer
can possibly detect the variation of such difference (\ref{de0}), by measuring the 
Casimir effect at different altitude $r_1$ and $r_2$, 
\begin{equation}
\delta {\cal E}(r_2)|_{\rm Casimir}- \delta {\cal E}(r_1)|_{\rm Casimir},
\label{variation}
\end{equation}
where $\delta {\cal E}(r)_{\rm Casimir}(r_2)$ is the difference of vacuum energies in the presence 
and absence of two conducting plates. Such a gravitational effect (\ref{variation}) on the vacuum 
energy seems too small to be seen. 

However, in a gravitational collapsing process approaching to the horizon of a black hole, with respect to
a static observer located at $R(t)$, where is the collapsing shell and gravitational field strongly 
and rapidly varies, the difference $\delta {\cal E}(r)$ 
(\ref{ddem}) could be very large and the vacuum state gains a large amount of 
gravitational energy. 
 
\section{Gravitational collapsing}\label{collapsing}

\begin{figure}[th]
\begin{center}
\includegraphics[]{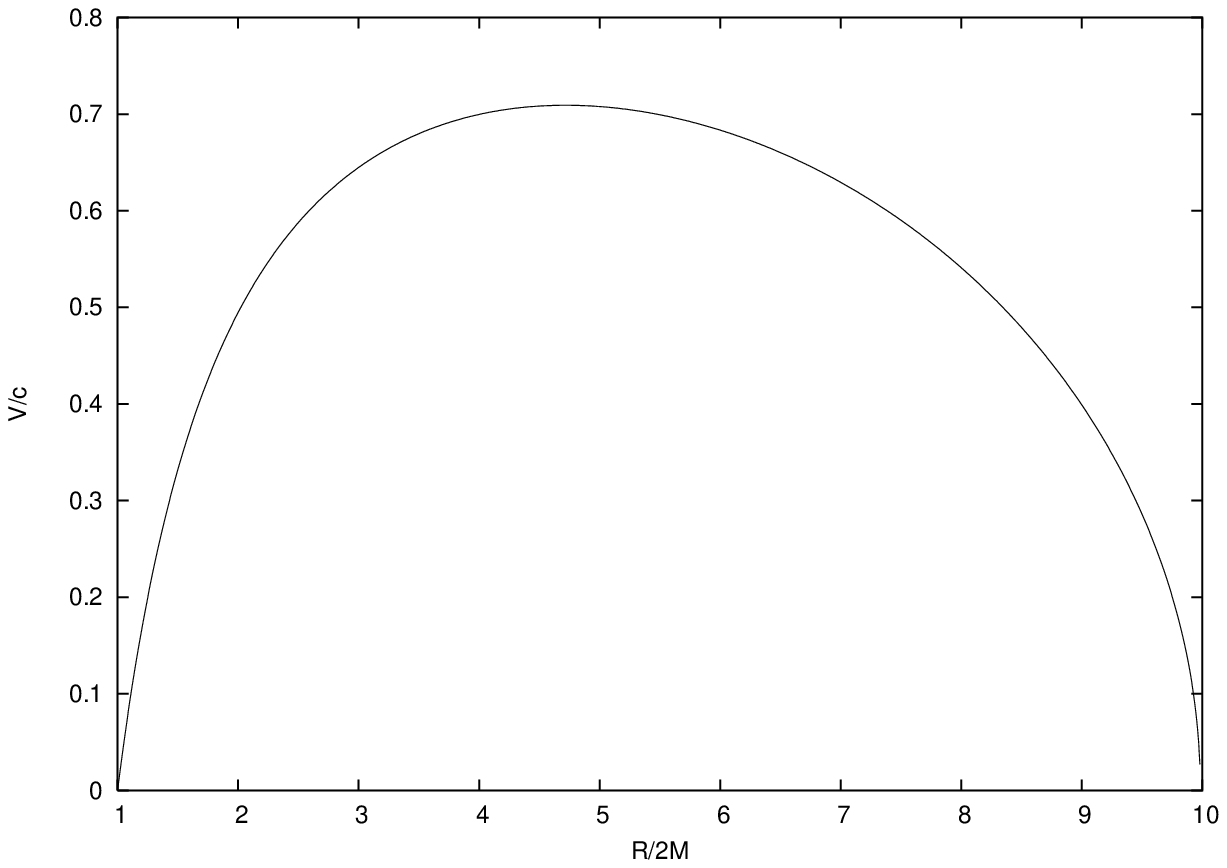}
\end{center}
\caption{The velocity of gravitational collapsing shell in the unit of $c$ as a function of radius $R$ 
in the unit of $2M$.}%
\label{fvelocity}%
\end{figure}

In order to be able to analytically study the vacuum-energy gain and in particular vacuum decay leading to 
photon productions in the following sections, we simplify the dynamical process of a gravitational collapse 
and adopt a model with the following approximations:
\begin{itemize}
\item
gravitational collapsing of massive and spherical shell that is infinitesimally thin;
\item
exactly spherical symmetry in gravitational collapsing;
\item
stationary Schwarzschild geometry in the external spacetime ${\cal S}$ outside the massive shell;
\item
flat geometry in the internal spacetime $\bar {\cal S}$ inside the massive shell.
\label{gmodel}  
\end{itemize}
This spherical shell has a total mass-energy $M$ and rest mass $M_0$. In this model, the 
solution to the Einstein equations was studied in refs.\cite{gcollapsing} and     
such a gravitational collapse process can be described by the equation,
\begin{eqnarray}
{\delta R\over\delta t} &=&{g(R)\sqrt{h^2(R)-g(R)}\over h(R)},\label{coll}\\
h(R)&=& \Gamma-{2M\over4R}{1\over\Gamma},
\nonumber
\end{eqnarray}
where $V(t)\equiv\delta R/\delta t$, in the unit of $c$, is the collapsing velocity in 
the opposite radial direction, and the collapsing parameter $\Gamma\equiv M/M_0$. 
The solution of this equation $R=R(t)$ is the radial 
location of the collapsing shell at the moment $t$. At the moment $t_\circ$ when the collapsing 
velocity $V(t_\circ)=0$, the shell, that is at rest and starts to collapse, is located at the 
radial position:
\begin{equation}
R_\circ(t_\circ)={1\over4}{2M\over\Gamma (1-\Gamma)}.
\label{rest}
\end{equation}
We chose $M=10M_\odot$ and $\Gamma=0.0257$, Eq.(\ref{rest}) gives $R_\circ(t_\circ)=10(2M)$. 
In Fig.(\ref{fvelocity}), we plot the collapsing velocity $V(t)$ as a function of $\bar R\equiv R/2M$, 
shell's radius in unit of $2M$. 
We find that the collapsing process $R(t)$ for $R_\circ(t_\circ)\ge R(t)>2M$ undergoes almost 
in the speed of light, however, it becomes very slow when it approaches the horizon: $R(t)\rightarrow 2M$. 
The collapsing process takes about $0.004$ seconds. 

When such collapsing shell $R(t)$ sweeps inwards $\delta R$ in the time interval $\delta t$, 
with respect to the static observer located at $R(t)$, the vacuum state changes from 
$|\bar 0, {\rm in}\rangle$ (\ref{vd1}) to 
$|\tilde 0,{\rm out}\rangle$ (\ref{vd}). The amplitude of vacuum state transition is given by 
Eq.(\ref{zr}), correspondingly the variations of energy-momentum densities are given by 
Eqs.(\ref{tem},\ref{tem1}). The vacuum-energy variation is given by Eq.(\ref{ddem}),
\begin{eqnarray}
\delta{\cal E}(R) &=& (g^{1/2}(R)-1){\cal E}_{\rm in}(R)\nonumber\\
&=& (g^{1/2}(R)-1)\left({5\over6}\right)
{\Lambda^2\over 2\pi}\delta R,
\label{oddem}
\end{eqnarray}
indicating that vacuum state $|\tilde 0, {\rm out}\rangle$ (\ref{vd}) gains gravitational energy with 
respect to $|\bar 0, {\rm in}\rangle$ (\ref{vd1}). By using the collapsing equation (\ref{coll}), 
we obtain the rate of vacuum-energy variation per unit of time:
\begin{equation}
{\delta{\cal E}(R)\over \delta t}= \left({5\over6}\right){\Lambda^2_p\over 2\pi}
(1-g^{1/2}(R)){\delta R\over\delta t},
\label{tddem}
\end{equation}
where and henceforth we take the absolute value of the collapsing velocity 
$V=\delta R/\delta t$. This is the rate of the vacuum states gaining gravitational 
energy in gravitational collapsing process.

In order to see the numbers of this rate in Eq.(\ref{tddem}), we take the ultraviolet 
cutoff $\Lambda=\Lambda_p$ and convert the natural unit 
($\Lambda_p=1$) into: $\Lambda^2_p= {\Lambda_p\over t_p}=1.95\cdot 10^{16}
{\rm ergs}/5.4\cdot  10^{-44}{\rm sec}= 3.6\cdot 10^{59}$egrs/sec,  
\begin{equation}
{\delta{\cal E}(R)\over \delta t}= 4.78\cdot 10^{58}
(1-g^{1/2}(R)){\delta R\over\delta t}\left({\rm egrs\over \rm sec}\right).
\label{tddem'}
\end{equation}
It is worthwhile to point out that the number $3.6\cdot 10^{59}$egrs/sec
in the rate of the vacuum-energy gain is completely 
determined by natural constants, independently of any free parameters.

Eqs.(\ref{tddem'}) completely determines the rate of vacuum-energy gain 
$\delta {\cal E}(R)/\delta t$ in the spherical shell $4\pi R^2\delta R$ that the 
collapsing shell sweeps inwards in the time interval $\delta t$. We plot the rate 
of vacuum-energy variation (gain) $\delta {\cal E}(R)/\delta t$ in Fig.(\ref{frate}). 
The result shows that the rate $\delta {\cal E}(R)/\delta t$ rapidly increases to 
$10^{57}$erg/sec, as the radius $R(t)$ of the collapsing shell 
moves, almost in the speed of light, inwards to the horizon. Whereas, in the 
vicinity of the horizon, the collapsing process becomes slow 
and the rate $\delta {\cal E}(R)/\delta t$ decreases and goes to zero.

The total amount of vacuum-energy gain from gravitational field at the end of 
gravitational collapse is given by integrating Eq.(\ref{oddem}):
\begin{eqnarray}
{\cal E}_{\rm total}&=&\int_R \delta {\cal E}(R)=\left({5\over6}\right)
{\Lambda^2_p\over 2\pi} 
\int_{R_\circ}^{2M}(g^{1/2}(R)-1)
\delta R \nonumber\\
& =& \left({5\over6}\right)
{2M\over 2\pi}\int_{10}^1(g^{1/2}(\bar R)-1)\delta \bar R\simeq 0.1 M 
\label{ttddem}
\end{eqnarray}
where $R_\circ =10(2M)$.  
The maximum variation of gravitational energy is $M/2$ in the collapse 
process, which can be derived from differentiating the gravitational potential 
($-M/ r$) from $r\sim\infty$ to $r=2M$. 
   
Due to this vacuum-energy gain $\delta {\cal E}(R)$ (\ref{oddem}), 
vacuum states become energetically unstable, have to spontaneously undergo a quantum
transition to lower energy states via quantum-field fluctuations, which leads to particle
productions. As a consequence, the vacuum-energy $\delta {\cal E}(R)$ (\ref{oddem}) 
gained from gravitational field must be released and deposited in the region 
from $r=2M$ to $r=R_\circ$.
 
\section{Energy release and photon productions}\label{production}

Which process of 
quantum transition releases this vacuum-energy gain $\delta {\cal E}(R)$ (\ref{oddem})? 
One of possibilities is spontaneous photon emission, analogously to the
spontaneous photon emission taking place in the atomic physics. In ref.\cite{xueshort2003}, we mentioned 
the possibility of such a spontaneous 
photon emission can be induced by the four-photon interacting vertex in the 
Quantum Electromagnetic Dynamics (QED). Yet, we have not been able to calculate
the rate of such spontaneous photon-emission, since we are studying quantum scalar fields, 
instead of the QED, in curved spacetime. In this section, we try 
to compute the quantum transition amplitude between the initial and final vacuum states, corresponding to 
before and after gravitational field is turned on, at each step of gravitational collapsing process. 
This quantum transition amplitude is
related to the probability of spontaneous particle (``photon'') productions.
 
First, we define the invariant scalar product of initial vacuum state $\phi_{\rm in}(x)$ and final vacuum state
$\phi_{\rm out}(x)$ (see Eq.(3.28) in the book by Birrell and Davies\cite{book}):
\begin{eqnarray}
(\phi_{\rm out},\phi_{\rm in})&=&-i\int_\Sigma \phi_{\rm out}(x)\bar\partial_\mu\phi_{\rm in}^*(x) d\Sigma^\mu
\label{product}\\
\phi_{\rm out}(x)\bar\partial_\mu\phi_{\rm in}^*(x)&=&\phi_{\rm out}(x)\partial_\mu\phi_{\rm in}^*(x)-\partial_\mu [\phi_{\rm out}(x)]\phi_{\rm in}^*(x)
\nonumber 
\end{eqnarray}
where $d\Sigma^\mu=n^\mu d\Sigma$, with a future-directed unit orthogonal to the spacelike 
hypersurface $\Sigma$ and $d\Sigma$ is the volume element in $\Sigma$. Since the value of 
$(\phi_{\rm out},\phi_{\rm in})$ is independent of $\Sigma$, we rewrite Eq.(\ref{product}) as,  
\begin{equation}
(\phi_{\rm out},\phi_{\rm in})=-i\int_{\Sigma_t}  d\Sigma^t \phi_{\rm out}(x)\bar\partial_t\phi_{\rm in}^*(x)
\label{product1}
\end{equation}
where $\Sigma_t$ is the spacelike hypersurface for $t=$constance and its element  
$d\Sigma^t=n^t d\Sigma$ is given in Eq.(\ref{vector}).

With respect to the rest observer ${\cal O}$ located at the radial position $r=R(t)$ and 
at the moment $t$, where and when the collapsing shell $R(t)$ 
sweeps inwards, the gravitational field is turned on for the time interval $\delta t$. This
time scale $\delta t$ is determined by the gravitational collapsing process $\delta R$ (\ref{coll}). 
In this time interval $\delta t$, the vacuum state changes from the 
initial vacuum state $\phi_{\rm in}(x)$: 
\begin{equation}
\phi_{\rm in}(x)=
\left\{
\begin{array} [c] {c} {1\over\sqrt{2\omega_0}}e^{-i\omega_0 t_0}\bar R_{\omega_0l_0}(r) Y_{l_0m_0}(\Omega),\hskip0.2cm r<R+{\delta R\over2}\\
{1\over\sqrt{2\omega}}e^{-i\omega t} R_{\omega l}(r) Y_{lm}(\Omega),\hskip1.2cm r>R+{\delta R\over2}
\end{array}
\right.
\label{phii}
\end{equation}
to the final vacuum state $\phi_{\rm out}(x)$:
\begin{equation}
\phi_{\rm out}(x)=\left\{\begin{array} [c] {c} {1\over\sqrt{2\omega_0}}e^{-i\omega_0 t_0}\bar R_{\omega_0l_0}(r) Y_{l_0m_0}(\Omega),\hskip0.2cm r<R-{\delta R\over2}\\
{1\over\sqrt{2\omega}}e^{-i\omega t} R_{\omega l}(r) Y_{lm}(\Omega),\hskip1.2cm r>R-{\delta R\over2}.
\end{array}\right.
\label{phif}
\end{equation}
We assume that the time scale of quantum-field transition 
from the initial vacuum state $\phi_{\rm in}(x)$ to the final $\phi_{\rm out}(x)$ induced by 
the variation of gravitational field is very much shorter than the gravitational collapsing 
time-scale $\delta t$, so that the gravitational field is adiabatically turned on at $r=R(t)$, 
the initial vacuum state $\phi_{\rm in}(x)$ and final vacuum state $\phi_{\rm out}(x)$ 
are considered as their asymptotic eigenstates ($\omega_0,l_0,m_0$) and ($\omega,l,m$) 
respectively.

With these initial state $\phi_{\rm in}(x)$ and final states $\phi_{\rm out}(x)$, Eq.(\ref{product1}) 
gives the vacuum to vacuum transition amplitudes, when the gravitational field is turned on
at $r=R(t)$. These vacuum to vacuum transition amplitudes are just the 
Bogolubov coefficients:
\begin{equation}
\alpha_{ij}=(\phi_{\rm out},\phi_{\rm in});\hskip0.3cm \beta_{ij}=-(\phi_{\rm out},\phi^*_{\rm in}),
\label{bogo}
\end{equation} 
where $|\beta_{ij}|^2$ describes the probability of particle productions.
Using $\phi_{\rm in}$ (\ref{phii}) and $\phi_{\rm out}$ (\ref{phif}), we compute the transition amplitude 
$\beta_{ij}=(\phi_{\rm out},\phi^*_{\rm in})$
\begin{eqnarray}
(\phi_{\rm out},\phi^*_{\rm in})&=&
\int_{[0,R-\delta R/2]}d\Sigma^t(\omega_0-\omega_0) \phi_{\rm out}\phi_{\rm in}\nonumber\\
&+&
\int_{[R-\delta R/2,R+\delta R/2]}d\Sigma^t(\omega -\omega_0) \phi_{\rm out}\phi_{\rm in}\nonumber\\
&+&\int_{[R+\delta R/2,\infty]} d\Sigma^t (\omega-\omega)\phi_{\rm out}\phi_{\rm in}\nonumber\\
&=&\int_{[R-\delta R/2,R+\delta R/2]}d\Sigma^t (\omega-\omega_0)\phi_{\rm out}\phi_{\rm in},
\label{trans}
\end{eqnarray}
where $[A,B]$ indicates the integration zone of $\Sigma^t$ in the radial direction, 
and in the last line of equation
\begin{eqnarray}
\phi_{\rm out}\phi_{\rm in} &=&{1\over\sqrt{2\omega}}e^{-i\omega t} R_{\omega l}(r) Y_{lm}(\Omega)\nonumber\\
&\cdot&{1\over\sqrt{2\omega_0}}e^{-i\omega_0 t_0}\bar R_{\omega_0l_0}(r) Y_{l_0m_0}(\Omega),
\label{crossing}
\end{eqnarray}
where $\omega=g^{1/2}(r)\omega_0$, 

Summing over initial states, we obtain the probability of particle productions in 
final states within the energy interval $(\omega, \omega+d\omega)$, 
\begin{eqnarray}
{dN\over d\omega}&=&\sum_{\omega_0}|\beta_{ij}|^2\nonumber\\
&=&\int_{R-\delta R/2}^{R+\delta R/2}r^2dr
{(1-g^{1/2}(r))^2\over 4g^{1/2}(r)}|R_{\omega l}(r)|^2\nonumber\\
&\simeq &{(1-g^{1/2}(R))^2\over 4g^{1/2}(R)}
\delta R R^2|R_{\omega l}(R)|^2,
\label{bc}
\end{eqnarray}
where $l=l_0$ and $m=m_0$. To derive Eq.(\ref{bc}) we use the orthogonality and closure 
relations of eigenfunctions $R_{\omega_0l_0}(r) Y_{l_0m_0}(\Omega)$: 
\begin{eqnarray}
&(1)&\int d\Omega Y^*_{lm}(\Omega)Y_{l_0m_0}(\Omega)=\delta_{ll_0}\delta_{mm_0};\nonumber\\
&(2)&\sum_{\omega_0l_0m_0}\left(\bar R_{\omega_0l_0}(r) Y_{l_0m_0}(\Omega)\right)^*
\bar R_{\omega_0l_0}(r') Y_{l_0m_0}(\Omega')\nonumber\\
&=&{1\over r^2}\delta(r-r')\delta^2(\Omega-\Omega');
\nonumber\\
&(3)&\int_{\Sigma_t}d\Sigma^t {1\over r^2}\delta(r-r')\delta^2(\Omega-\Omega')f(r,\Omega)
=f(r',\Omega').\nonumber
\end{eqnarray}
Using the rate $\delta R/\delta t$ given by gravitational collapsing equation (\ref{coll}), 
we obtain the rate of the particle productions, 
\begin{equation}
{dN\over dt d\omega}\simeq {(1-g^{1/2}(R))^2\over 4g^{1/2}(R)}
{\delta R \over \delta t}R^2|R_{\omega l}(R)|^2.
\label{bc1}
\end{equation}
This equation gives the rate and spectrum of particle productions, corresponding to the 
vacuum-energy variation $\delta{\cal E}(R)/\delta t$ (\ref{tddem}) in the simplest 
model of gravitational collapsing shell, as described in the beginning of 
section (\ref{collapsing}). 

\begin{figure}[th]
\begin{center}
\includegraphics[]{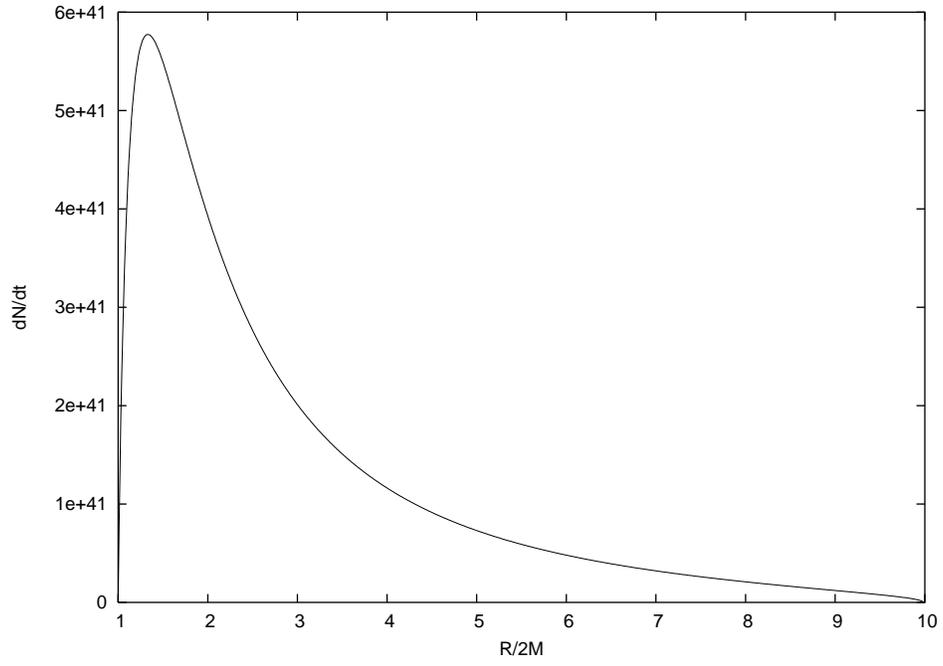}
\end{center}
\caption{The rate of particle creations (1/sec) as a function of radius $R$ in the unit of $2M$.}%
\label{fratenumber}%
\end{figure}

In order to have an idea of the number of particle creations in a second, we
approximately use the continuity of functions $\bar R_{\omega_0 l}(R)$ and $R_{\omega l}(R)$,
at $r=R$
\begin{equation}
R_{\omega l}(R)\simeq \bar R_{\omega_0 l}(R),
\label{continue}
\end{equation}
where is infinitesimally thin shell in gravitational collapsing.
As a result, we have 
\begin{equation}
{dN\over dt d\omega} \simeq {(1-g^{1/2}(R))^2\over 4g^{1/2}(R)}
{\delta R \over \delta t}R^2|\bar R_{\omega_0 l}(R)|^2,
\label{bc3}
\end{equation}
where $\bar R_{\omega_0 l}=2\omega_0 j_l(\omega_0 R)$. Using the relation $\omega=g^{1/2}(R)\omega_0$
and integrating $\omega_0$ over $[0,\Lambda]$ in Eq.(\ref{bc3}) for
the S-wave $(l=0)$, we obtain,
\begin{eqnarray}
{dN\over dt } &\simeq& {\Lambda\over2} (1-g^{1/2}(R))^2
{\delta R \over \delta t},\label{nbc3}\\
&=&\! 9.26\cdot 10^{42}(1\!-\!g^{1/2}(\bar R))^2
{\delta R \over \delta t}\left({1\over\rm sec.}\right),
\nonumber
\end{eqnarray}
where we take $\Lambda=\Lambda_p$ again. We plot the rate of particle creations 
$dN/ dt$ in terms of $\bar R=R/2M$ in Fig.(\ref{fratenumber}), which shows 
that the rate $\delta N/\delta t$ rapidly increases to 
$10^{43}$/sec, as the radius $R(t)$ of the collapsing shell 
moves, almost in the speed of light, inwards to the horizon. Whereas, in the 
vicinity of the horizon, the collapsing process becomes slow 
and the rate $\delta N/\delta t$ decreases and goes to zero.
The total number of particles created in the collapse process is given by integrating 
Eq.(\ref{nbc3}),
\begin{eqnarray}
N &=& {\Lambda_p\over2} (2M)\int_1^{10} 
(1-g^{1/2}(\bar R))^2\delta\bar R,\nonumber\\
&=&0.361\Lambda_p \mu M_\odot=6.69\cdot 10^{38}.
\label{tnbc3}
\end{eqnarray}

\begin{figure}[th]
\begin{center}
\includegraphics[]{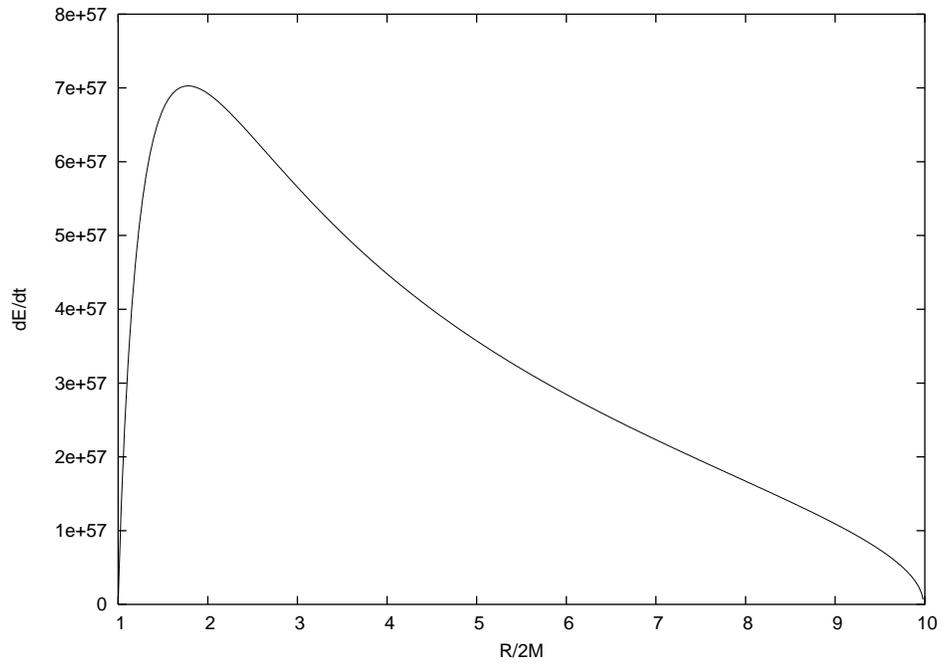}
\end{center}
\caption{The rate of vacuum-energy gain (ergs/sec) as a function of radius $R$ in the unit of $2M$.}%
\label{frate}%
\end{figure}

As shown in Figs.(\ref{frate}) and (\ref{fratenumber}), the rate of vacuum-energy gain,
and the rate of particle creations are very large,
as the collapsing process approaching to the formation of black hole's horizon $R=2M$.
The total energy output (\ref{ttddem}) and number (\ref{tnbc3}) of particles created are 
enormous.
These qualitatively agree to the characteristics of energetic sources for gamma ray bursts.
It is indeed interesting that the Planck scale $\Lambda_p$ as the ultraviolet 
cutoff in our proposal neutrally gives rise to the characteristics of gamma ray bursts,
instead of depending on an arbitrary energy scale.

The energy of photons spontaneously emitted can be larger than the energy threshold 
$2m_e$, so that electron and positron pairs are produced. These pairs, on the other 
hand, annihilate into two photons. As a consequence, a dense and energetic plasma of 
photons, electron and position pairs, called ``{\it dyadosphere}'' \cite{ruffini} 
or ``{\it fireball}'' in literatures\cite{piran}, could be formed. The energy and 
particle-number densities of this ``{\it dyadosphere}'' can be respectively obtained by 
Eq.(\ref{tddem'}) and Eq.(\ref{nbc3}), see Figs.(\ref{frate},\ref{fratenumber}). 
The total energy and particle-number of ``{\it dyadosphere}'' are given by 
Eq.(\ref{ttddem}) and Eq.(\ref{tnbc3}). 

\section{Comparison with Sonoluminiescence}\label{sololu}

Sonoluminiescence is the phenomenon of the intense flashes of light emission by the pulsations of 
a gas bubble driven by sound-wave in fluid\cite{sololuexp}. Such experiments deal with the pulsations 
of bubbles of air in water, driven by a sound wave of frequence of 
20-30 KHz. During the expanding phase, the bubble radius reaches maximum 
of order $R\sim 4.5\mu m$, followed by a rapid collapse down to a minimum radius of order $R\sim 0.5\mu m$.
The photons are emitted, having a ``quasi-thermal'' spectrum with a 
``temperature'' of several tens of thousands of degrees Kelvin. There are about $10^6$ photons 
emitted per flash, and the time-averaged total power emitted is between 30 and 100 mW. The photons
appear to be emitted a very tiny spatio-temporal region: Estimated flash widths vary from 
less than 35 ps to more than 380ps depending on the gas in the bubble\cite{sololuexp}.

The fundamental mechanism of such photon emissions in this phenomenon is still very 
controversial\cite{sololuth}. We do not want to enter these controversial discussions in this 
article. In this section, we attempt to 
briefly discuss the Schwinger proposal\cite{swsolo} in the connection 
with our study of large photon productions in a gravitational collapse. Schwinger considered this 
phenomenon of photon emissions as the Casimir energy (vacuum energy) $E$ releasing, due to the 
variation of the Casimir energy when a very rapid collapse of dielectric material into 
a vacuum takes place, 
\begin{equation}
E=-\int{d\vec r d\vec k\over(2\pi)^3}{1\over2}|\vec k|\left(1-{1\over \sqrt{\epsilon(\vec r)}} \right),
\label{sche}
\end{equation}
where $\epsilon(\vec r)$ is dielectric constant and $|k|$ the vacuum energy of zero-point fluctuation mode. 
The total excess energy is
\begin{equation}
|E|={1\over12\pi}R^3K^4\left(1-{1\over \sqrt{\epsilon}} \right),
\label{sche1}
\end{equation}
for a slow varying dielectric constant $\epsilon$, where $R$ is bubble's radius and $K$ is a cut-off wavenumber.
The dielectric constant $\epsilon\rightarrow 1$, with respect to the high-energy modes above the cutoff $K$.  
If this Casimir energy releasing is completely in form of photon emissions, one identifies the average 
number of photon emissions as, 
\begin{eqnarray}
N&=&\int{d\vec r d\vec k\over(2\pi)^3}{1\over2}\left(\sqrt{\epsilon}-1 \right),\nonumber\\
&=& {1\over9\pi}R^3K^3\left(\sqrt{\epsilon}-1 \right).
\label{sche2}
\end{eqnarray}
The cut-off wavenumber $K\sim 10^5$cm$^{-1}$ within the ultraviolet region, the energy-budget (\ref{sche1}) and
the number of photon emissions $N$ (\ref{sche2}) agree with the experiments, although the spectrum of 
Sonoluminiescence does not extend to the ultraviolet region. However, 
Schwinger neither explicitly worked out the mechanism of photon productions nor computed the 
rate of photon productions in the dynamical circumstance that the vacuum-energy variation is very rapid, 
for very rapid collapse of bubble. On the basis of this proposal, there are many further 
studies in the literatures\cite{sololuth}, concerning on mechanism of photon productions and other 
relevant aspects relating the Schwinger proposal to the experiment of Sonoluminiescence.      
We want compare our proposal presented for gamma ray bursts in this article with the 
Schwinger proposal for the phenomenon of Sonoluminiescence. Regarding the variation 
of the vacuum energy, we find that vacuum-energy variation (\ref{ddem}) due to turning on 
gravitational field is similar to vacuum energy 
variation (\ref{sche}) due to changing the dielectric constant. 
The variation of vacuum energies in both equations is negative ($\epsilon>1$), implying the vacuum state 
gains energy in both cases. In Eq.(\ref{ddem}), the vacuum state gains the gravitational energy. 
While, in Eq.(\ref{sche}), the vacuum state gains the sound-wave energy.
In both cases, the variation of vacuum energy is very rapid, because collapsing processes 
driven by either gravitational field or sound-wave are very rapid. The collapsing velocity 
$\dot R\simeq c$ in the gravitational collapsing case and $\dot R\simeq 4$March in the
Sonoluminiescence case. The cutoff wavenumber $K$ is a real physical cutoff 
of its own right that Eqs.(\ref{sche1},\ref{sche2}) make physical scenes up to this cutoff. Analogously,
The scale $\Lambda$ is a real physical cutoff of its own right that Eqs.(\ref{energy1},\ref{ddem}) 
take into account the total variation of vacuum energy, attributed to turning on an external gravitational 
field. It is not an artificial cutoff introduced for regulating calculations of divergent terms, and then removed 
in renormalizable theories.  Although Schwinger did not explicitly work out the mechanism of 
photon productions in the phenomenon of Sonoluminiescence, the basic idea for vacuum-energy variation and 
photon productions is similar to that we propose in ref.\cite{xueshort2003} and this article: 
the vacuum state gains (sound-wave/gravitational) energy and becomes unstable and must decay to the 
lowest energy state, releasing the (sound-wave/gravitational) energy it gains. 

\section{Contrastion with the Hawking effect}\label{hawking}

It is important to differentiate the Hawking radiation from the effect discussed in 
this article. It is clear that both effects are attributed to an external gravitational field
interacting with virtual particles in the vacuum. However, they are very different, not only 
in the phenomena of their appearances, but also dynamics of their origins. 

First, we see the 
aspect of phenomenon. The Hawking radiation is black-body radiation from a thermal bath of 
the temperature $1/8\pi M$ determined by the scale $M$. Particle creations leading to the Hawking 
radiation do not depend on gravitational collapse processes, or in the other words, 
the Hawking radiation can be created by an external {\it static} gravitational field. While, particle
creations discussed in this article do not have a black-body spectrum and the energy-scale 
of particle creations processes is a ultraviolet cutoff $\Lambda$. 
Such particle creations strongly depend on the gravitational collapsing processes, or in the 
other words, particle creations discussed in this article cannot occur in an external 
static gravitational field. 

Second, we discuss the aspect of dynamics. There are many elegant scenarios interpreting 
the origin of the Hawking radiation around the horizon of a black hole. 
In the recent article\cite{xueNP2003}, based on the context of quantum field theories for particle and 
antiparticle creations in an external gravitational field, 
author presented a scenario for understanding the origin of quantum radiation of the Hawking type when 
gravitational field is present. Since the general formulation of quantum scalar-field 
theory discussed in sections (\ref{general}) and (\ref{spectrum}) is similar to that 
in ref.\cite{xueNP2003}, we adopt our scenario to contrast two
different dynamical origins of the Hawking radiation and particle creations discussed in this article.        

As discussed in the ref.\cite{xueNP2003}, a gravitational field polarizes the vacuum: 
quantum field fluctuations (creation and annihilation) of virtual particles and antiparticles (positive and 
negative energy states) in the vacuum are ``aligned'' by an external gravitational field. 
By this gravitational polarization effect, the vacuum gains gravitational energy.
This energy-gain reduces the energy-mass gap that is a barrier, preventing virtual particles in the 
vacuum from tunneling and creating particles. As a result, the probability of such quantum 
tunneling is increase. This causes the vacuum decay and creations 
of particles and antiparticles. This quantum tunneling effect is quantitatively described 
by the imaginary term, $i2M/r^2$
in Eq.(\ref{ipro0}) and imaginary effective action. This is rather analogous to the QED vacuum 
in the presence of 
an external electric field. The quantum-filed fluctuations of charged virtual particles 
are polarized by the electric field. Particles and antiparticles are created by the Schwinger 
mechanism\cite{sw}, when electric field strength is strong enough to overcome energy-mass gap $\sim 2m_e$. 
The thermal nature of quantum radiation of the Hawking type is due to the CTP invariance in 
the processes of particle creations and annihilations.
The temperature of thermal radiation is determined by the vacuum-energy gain when gravitational field 
polarizes the vacuum. The reason why the temperature (or vacuum-energy gain) is the order of $M^{-1}$ has 
been discussed in section (\ref{scale}).   

Instead, as discussed in previous sections, the dynamics of particle creations we discussed in this
article is very similar to the dynamics of mechanism that Schwinger discussed for 
Sonoluminiescence. It is also rather similar to the dynamics of photon-creations of the dynamical
Casimir effect\cite{dyacasimir}. For the reasons that gravitational field interacting with virtual 
particles in the vacuum and the variation of gravitational field in the collapse process,
the vacuum gains gravitational energy and the large variation of vacuum energy occurs in a very 
short time and small space. Such a large vacuum-energy gain makes the vacuum state to be energetically 
unstable. Via quantum-field fluctuations, unstable vacuum state has to transit to lower 
energetical vacuum-state. Such vacuum to vacuum transition releases the 
gravitational energy, that the vacuum state gains, by spontaneous
photons emissions. On the basis of its dynamical origin, this process of photon productions clearly 
does not take place in an external static gravitational field, very differently from the 
thermal radiation of the Hawking effect. It can be seen from computations in previous sections that 
this process is mainly contributed by the variation of terms 
$-g^{tt}\omega^2-g^{rr}k_r^2$ in Eq.(\ref{ipro0}), during a gravitational collapse. This 
contrasts with the imaginary term $i2M/r^2$ in Eq.(\ref{ipro0}), describing quantum tunneling 
process for the Hawking effect in a static gravitational field. The energy scales of 
two processes are very different, which have been discussed in section (\ref{scale}).
    
\section{Some Remarks }\label{discussion}

The research of our proposal is at a preliminary step. We adopt the action (\ref{action}) for scalar fields, 
rather than the vectorial field of electrodynamics dynamics (QED) in curved spacetime. The notion of 
photon productions in the title and text of the present article should be replaced by
``photon'' productions. In appendix A, we make an approximation in computing non S-wave 
contributions. We would like to consider the results (\ref{fl0},\ref{sgl1}) as the S-wave contribution only.  
In addition, we adopt an approximate model (infinitesimally thin shell) for describing the process 
of gravitational collapse. Learning a controversy\cite{sololuth} on the ultraviolet cutoff introduced 
in the Schwinger proposal for Sonoluminiescence, we need to further strengthen our discussions and 
arguments on the ultraviolet cutoff $\Lambda$ and its value in our proposal. As discussed in section 
(\ref{sololu}), Schwinger introduced the ultraviolet cutoff $K$ to agree with the energy 
budget and particle number of Sonoluminiescence, although the spectrum of Sonoluminiescence 
does not extend to this ultraviolet region. Analogously, the ultraviolet cutoff $\Lambda$ 
at the Planck scale is introduced in our proposal to be consistent with the energy budget and 
particle number of gamma ray bursts in sections (\ref{collapsing},\ref{production}), 
although the spectrum of gamma ray bursts is $O$(MeV). From the spectrum of particle creations 
Eqs.(\ref{bc1}) or (\ref{bc3}), 
we find that particle creations are dominated in the low-energy region, since the spectrum is 
approximately related to the function $\sin^2(\omega R)/(\omega R)^2$ for a given value of 
collapsing radius $R$. We speculate that high-energy particles should lead to multiparticle
productions, and total number of particles produced is much larger than $N$ (\ref{tnbc3}).  
We are still far from a complete understanding of our proposal for gamma ray bursts. 
Nevertheless, it is highly 
deserved to study the proposal presented in this article in connection with the 
origin of gamma ray bursts.

It is a good analogy to compare our proposal for the origin of gamma ray bursts with 
Schwinger proposal for the origin of Sonoluminiescence. Further experimental and 
theoretical studies on the Schwinger proposal for Sonoluminiescence will definitely 
help us to have a better understanding of the origin of gamma ray bursts in our proposal. 
Beside, experimental and theoretical studies on photon productions in the dynamical 
Casimir effect are essential
for us to further understand the mechanism of photon productions in our proposal. 

In literatures\cite{sologrb}, we find that the Schwinger idea for Sonoluminiescence has been applied
for explaining the origin of gamma ray bursts, on the basis of the variation of dielectric constant 
during a gravitational collapse. These scenarios seem interesting, in particular, in explaining the 
total energy budget of gamma ray bursts. Analogously, conducting electron gas is used as boundary 
conditions for computing the Casimir energy to discuss possible huge output of cosmic energy 
accounting for Quasars\cite{Sokolov}.

In future work, we expect to be able to study the quantum field theory of electrodynamics dynamics (QED) 
in curved spacetime and use more precise model describing the process of gravitational collapse, 
as well as elaborate calculations of vacuum-energy density and 
rate of gravitational energy releasing by spontaneous photon productions.

\section{Acknowledgments}\label{ack}
I am grateful to R.~Ruffini for bring me into this arena of physics and his continuous support. I thank to 
L.~Vitagliano for many discussions on the issues of observer and gravitational collapse.

\section{Appendix A}\label{computationl}

In this appendix, we calculate the non S-wave ($l\not=0$) contributions 
$\langle T^t_t\rangle_{\rm in}^{l\not=0}$
in Eq.(\ref{sl0}) and $\langle T^t_t\rangle_{\rm out}^{l\not=0}$ in Eq.(\ref{sgl}).

In Eq.(\ref{sl0}), the non S-wave ($l\not=0$) contributions 
$\langle T^t_t\rangle_{\rm in}^{l\not=0}$ is given by,
\begin{eqnarray}
\langle T^t_t\rangle_{\rm in}^{l\not=0} &=& 2{1\over 4\pi r^2}\sum_{l=1}^\infty (2l+1)
\int{dk_{r0}\over (2\pi)}\nonumber\\
&\cdot&\sqrt{k_{r0}^2+ {l(l+1)\over r^2}+ m^2},
\label{l}\\
&=&{\Gamma(-{\epsilon\over2})\over 4\pi^2 r^2}\sum_{l=1}^\infty (2l+1)
\left[{l(l+1)\over r^2}+ m^2\right]^{(1+\epsilon)},
\label{l1}
\end{eqnarray} 
where we use the formula (\ref{fk}) in Eq.(\ref{l1}). 
Using Eqs.(\ref{fk}) and (\ref{changing}), we compute 
$\langle T^t_t\rangle_{\rm out}^{l\not=0}$ in Eq.(\ref{sgl}),
\begin{eqnarray}
\langle T^t_t\rangle_{\rm out}^{l\not=0}&=&2{g^{1/2}\over 4\pi r^2}\sum_{l=1}^\infty 
(2l+1)\int{dk_r\over (2\pi)}\nonumber\\
&\cdot&\sqrt{gk_r^2\!+\! V_{l\not=0}\!-\!i{2M\over r^2}k_r},
\label{gl}\\
&=&{\Gamma(-{\epsilon\over2})\over 4\pi^2 r^2}
\left({l(l+1)\over r^2}+ m^2+Q^2\right)^{1+\epsilon}.
\label{gl5}
\end{eqnarray}
Based on $\langle T^t_t\rangle_{\rm in}^{l\not=0}$ (\ref{l1}) and 
$\langle T^t_t\rangle_{\rm out}^{l\not=0}$ (\ref{gl5}) for $\epsilon\rightarrow 0$ up 
to $O(2/\epsilon)$, we find
that the terms $m^2$ and $l(l+1)/r^2$ are canceled in the difference
$\langle T^t_t\rangle_{\rm diff}^{l\not=0}$ Eq.(\ref{tem1}). 

In the following 
calculations, we only keep up to the terms that are $O(2/\epsilon)$ in the limit of 
$\epsilon\rightarrow 0$. Thus, considering 
the difference $\langle T^t_t\rangle_{\rm diff}^{l\not=0}$ Eq.(\ref{tem1}), we write
Eq.(\ref{l}) and Eq.(\ref{gl}) as,
\begin{eqnarray}
\langle T^t_t\rangle_{\rm in}^{l\not=0} &=& 2{1\over 4\pi r^2}\sum_{l=1}^\infty (2l+1)
\int{dk_{r0}\over (2\pi)}\sqrt{k_{r0}^2},
\label{l'}\\
\langle T^t_t\rangle_{\rm out}^{l\not=0}&=&{2\over 4\pi r^2}\int{d\bar k_r\over (2\pi)}
\sum_{l=1}^\infty (2l+1)\sqrt{\bar k_r^2\!+\!Q^2}.
\label{gl1}
\end{eqnarray}
The summation over ``$l$'' in Eqs.(\ref{l'}) and (\ref{gl1}) is given by  
\begin{equation}
\sum_{l=1}^\infty (2l+1)=2\zeta(-1)+\zeta(0),
\label{l2}
\end{equation}
where $\zeta(n)$ is the Riemann zeta function 
\begin{equation}
\zeta(n)=\sum_{l=1}^\infty{1\over l^n},\hskip0.5cm 
n=0,1,2,3,\cdot\cdot\cdot.
\label{zeta}
\end{equation}
Resulted from summing over ``$l$'' in Eqs.(\ref{l'}) and 
(\ref{gl1}), $\zeta(-1)$ and $\zeta(0)$ are divergent.
As has been discussed, the transverse momenta $\vec k_\perp$ of transverse modes do not 
directly couple to the gravitational field, we discard these divergent terms, 
by assuming these divergent terms are independent of gravitational field $g(r)$. 
To eliminate these divergent terms, we use the reflection 
formula of analytic continued $\Gamma$- and $\zeta$-functions for complex variable 
$z=-n+\delta$ and $\delta\rightarrow 0$,
\begin{equation}
\Gamma(-{z\over2})\zeta(-z)=\pi^{-z-{1/2}}\Gamma({z+1\over2})\zeta(z+1).
\label{gx}
\end{equation}
As results,
We obtain
\begin{eqnarray}
\langle T^t_t\rangle_{\rm in}^{l\not=0} &=&-\left({1\over6}\right){2\over 4\pi r^2}
\int{dk_{r0}\over (2\pi)}\sqrt{k_{r0}^2},\nonumber\\
&\simeq &-\left({1\over6}\right) {1\over 4\pi r^2}\left({\Lambda^2\over 2\pi}\right);
\label{l6}\\
\langle T^t_t\rangle_{\rm out}^{l\not=0}&=&-\left({1\over6}\right){2\over 4\pi r^2}\int{d\bar k_r\over (2\pi)}
\sqrt{\bar k_r^2\!+\!Q^2},\nonumber\\
&\simeq& -\left({1\over6}\right){g\over 4\pi r^2}\left(
{\Lambda^2\over 2\pi}\right).
\label{sgl1n0}
\end{eqnarray}
In the second lines of these equations (\ref{l6}) and (\ref{sgl1n0}), we only keep the 
leading order $O(\Lambda^2)$, which is in accordance with the limit of 
$\epsilon\rightarrow 0$.   

It should be pointed that the definitions of $k_{r0}$ 
(\ref{kr01}) and $k^2_{r0}$ (\ref{kr201}) depend on the angular quantum number ``$l$'' 
and this implies that $\epsilon$ in Eqs.(\ref{l1}) is $l$-dependent. 
The exchanging the order of $k_{r0}$-integration and $l$-summation in 
Eqs.(\ref{l}-\ref{l1}) is not exact for a finite $\epsilon$, and the same problem for 
Eqs.(\ref{gl}-\ref{gl5})
On the other hand, the $l$-dependence in Eq.(\ref{l}) is dominated by the degeneracy term $2l+1$
and angular momentum term $l(l+1)$. Compared with these dominate terms, the $l$-dependence in 
$k_{r0}$ and $k_{r0}^2$ is very weak in terms of the radial wave-function $j_{l\omega_0}$, 
whose value is limited for $l\rightarrow\infty$. Thus, we neglect the $l$-dependence 
of $(k_{r0},k_{r0}^2)$ and $(k_{r},k_{r}^2)$ in computations. We consider this exchanging to be a 
good approximation for $\epsilon\rightarrow 0$. The S-wave results (\ref{fl0'},\ref{sgl10})
are free from these approximations.


\end{document}